\documentclass[12pt,prd,tightenlines,nofootinbib,showpacs,showkeys,setspace]{revtex4}
\usepackage{bm}
\usepackage{graphics}
\usepackage{rotating}
\usepackage{epsfig}
\usepackage{amsmath}
\usepackage{amsfonts}
\usepackage{setspace}
\usepackage{amssymb}

\begin{document}
\title{
Photonic production of the pair of $B_c$ mesons}
\author{\firstname{A.E.} \surname{Dorokhov}}
\affiliation{Joint Institute for Nuclear Research, BLTP, Moscow region,
Dubna, Russia}
\author{\firstname{R.N.} \surname{Faustov}}
\affiliation{Institute of Cybernetics and Informatics in Education, FRC CSC RAS, Moscow, Russia}
\author{\firstname{A.P.} \surname{Martynenko}}
\author{\firstname{F.A.} \surname{Martynenko}}
\affiliation{Samara National Research University, Samara, Russia}

\begin{abstract}
We study the pair production of $B_c$ mesons in the photon-photon interaction 
in the framework of perturbative quantum chromodynamics and the relativistic quark model. 
The production amplitudes of a pair of pseudoscalar and vector 
$B_c$ mesons are constructed in the nonrelativistic approximation and taking 
into account relativistic effects. Relativistic corrections related to the relative 
motion of heavy quarks in the production amplitude, as well as in the 
wave function of the bound state of heavy quarks, are taken into account. 
Analytical expressions are constructed for the relativistic differential and total 
cross sections for the pair $B_c$ meson production. Based on them, numerical 
values ​​of the production cross sections are obtained for various energies and scattering angles.
\end{abstract}



\maketitle

\section{Introduction}

Among the various reactions that are studied at modern elementary particle accelerators, 
the processes of the creation of bound states of heavy quarks $ (b, c) $ hold a special place. 
This is due to the fact that various theoretical methods have been developed 
to study the properties of heavy quarkoniums that have already proven their effectiveness. 
It cannot be said that the accuracy of theoretical calculations of the observed values 
​​for charmonium and others is very high. Nevertheless, different theoretical approaches 
allow us to calculate the mass spectrum, decay width, and production cross section 
of heavy quarkoniums so that the obtained results generally describe experimental data
\cite{gklt,brambilla2011,rqm1,pot,eq}. 
Thus, interest in the processes of production of heavy quarks is connected primarily 
with the ability to test the quantum chromodynamics and
the theory of bound states of particles. In this sense, 
interest to exclusive processes of pair production of heavy quarkoniums is enhanced 
by the fact that the effects of quark coupling are manifested here to a greater extent
\cite{eb,qiao,chao1,bodwin1,apm3,apm4}. Quarks are 
produced initially at short distances, almost free, and then diverge upon hadronization over long distances, at which the nonperturbative effects of their interaction become decisive.
As already shown by the studies of the production of a pair of charmoniums, such reactions 
make it possible to identify shortcomings in theoretical calculations of the production 
cross sections and to search for new mechanisms of interaction between quarks and gluons.

The study of various physical reactions in $\gamma\gamma$-interaction has always 
been part of the physical research program at electron-positron colliders. 
Such reactions arise as a result of the interaction of a cloud of virtual photons 
that are associated with accelerated charged particles. The transition from virtual 
to real photons was made possible due to the Compton back scattering of laser light
\cite{serbo1,serbo2}.
The production of high-energy real photons from light scattering by a 6-GeV electron beam, based 
on the Compton back scattering, was demonstrated in \cite{carlo}. Scattered photons acquired 
energies of hundreds of MeV and propagated mainly in the same direction 
as the electrons of the initial beam. A new round of interest in $\gamma\gamma$ interaction 
is currently connected primarily with the discovery of the Higgs boson, with the study 
of the $\gamma \gamma \to H $ process and the construction of the Higgs factory. 
The luminosity of the photon collider turns out to be related to the luminosity of the 
$ e^+ e^- $ collider by the relation: $ L_{\gamma \gamma} = k^2L_{ee} $, 
where the parameter $ k $ is defined as the fraction of electrons that produce 
the Compton photon. While $ e^+ e^- $ colliders have proven themselves in the 
study of various particle interactions, the creation of a photon collider together 
with the linear $ e^+ e^- $ collider, in which electron beams are converted 
to the photon beams, will allow conduct more effective research in a number of directions
\cite{telnov,badelek}. The study of different physical processes in which a pair 
of $ B_c $ mesons can be produced is important for understanding the possibilities 
of experimental detection of these particles.

The production of $B_c$ mesons is of particular interest in such reactions, 
since quarks of various flavours and masses arise in a bound state. Despite intensive 
experimental studies of $B_c$ mesons, information on them remains rather scarce. In fact, 
only the ground state of $B_c$ mesons ($0^-$) and the first excited state were observed. 
Single and paired production of $B_c$ mesons in $ e^+ e^- $ annihilation, 
$ pp $ interaction have already been studied for a long time both within the framework 
of the quark model and in nonrelativistic quantum chromodynamics. The first works on the single production of $B_c$ mesons in the $ \gamma \gamma $ interaction were performed in \cite{akl1,akl2} taking into account relativistic effects. 
The production of $B_c$ meson pair in $pp$ and $\gamma\gamma$ interaction was
estimated in \cite{baranov}. In our previous works \cite{apm1,apm2}, 
the pair production of $B_c$ mesons 
in electron-positron annihilation was studied, and we considered both one-photon 
and two-photon pair production mechanisms. In this paper, we consider the production 
of a pair of $B_c$ mesons in a collision of two real photons. 
The collision of high-energy photons, which can be obtained by the Compton back 
scattering of laser photons by high-energy electrons, can cause the exclusive production 
of a pair of heavy quarkoniums.
Since previous studies of the production of quarkoniums 
have revealed the important role of relativistic effects, we study the process 
$ \gamma + \gamma \to B_c^++ B_c^- $ both in the nonrelativistic approximation and taking 
into account relativistic corrections within our approach based on perturbative quantum 
chromodynamics and relativistic quark model \cite{apm3,apm1,apm2}.

\section{General formalism}

\begin{figure}[htbp]
\centering
\includegraphics[scale=0.5]{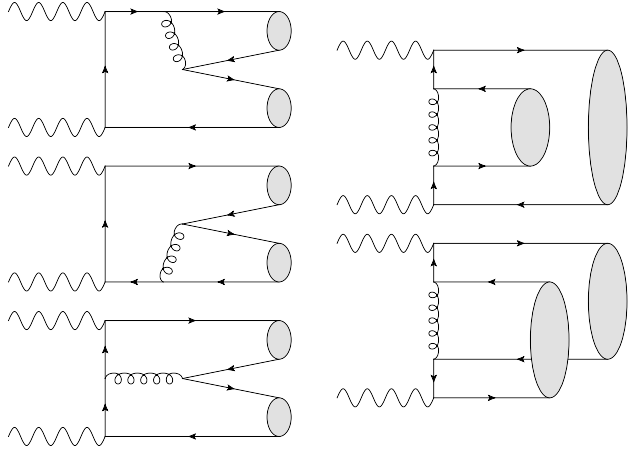}
\caption{The pair $B_c$-meson production amplitudes in $\gamma\gamma$ interaction.
Wavy lines show the real photons with four momenta $k_1$ and $k_2$.
Produced mesons $B^+_{c}$ and $B^{-}_{c}$
in the final state are indicated by a hatched oval.}
\label{fig1}
\end{figure}

The $B_c $ meson is the bound state of two heavy quark and anti-quark 
$ (\bar b, c) $ or $ (b, \bar c) $. At present, the ground state $B_c (0^-)$ with the mass 
$ M_{B_c} = 6274.9 \pm 0.8 $ MeV \cite{pdg} and two excited states 
$B_c(2^1S_0)^+$ with mass $6872.1\pm 1.3(stat)\pm 0.1(syst) \pm 0.8(B_c^+)$ MeV
and $B_c(2^3S_1)^+$ with mass $6841.2\pm 0.6(stat)\pm 0.1(syst) \pm 0.8(B_c^+)$ MeV
have been discovered \cite{pdg,aaij,avb}.
A new precision measurement of the $B_c^+$ meson mass is performed using proton-proton
collision data collected with the LHCb experiment in \cite{LHCb}.
For pair production of $B_c $ mesons, it is necessary to have a reaction in which 
two quark-antiquark pairs $ (c \bar c) $ and $ (b \bar b) $ would initially be produced. 
The reaction $ \gamma + \gamma \to B_c^++ B_c^- $ that we study is determined by the 20 Feynman amplitudes, some of which are shown in Fig.~\ref{fig1}. 
The Appendix A presents the entire set of production amplitudes of free quarks $(b,c)$
and antiquarks $(\bar b,\bar c)$ in the leading order of $\alpha_s$, necessary 
for the subsequent production of a pair of $B_c$ mesons.
These amplitudes are generated in the FeynArts package \cite{feynarts,feynarts1}.
The transition from the process of creating a pair 
of $B_c $ mesons in $ e^+ e^- $ annihilations to $ \gamma \gamma $ interaction, therefore, 
leads to a significant increase in the number of amplitudes and complicates the calculation of the pair production cross section.

The general expression for the total production amplitude of a pair of $B_c$ mesons 
can be written as a convolution of the production amplitude of two quarks $ (b, c) $, 
two antiquarks $ (\bar b, \bar c) $ in photon-photon collision and quasipotential relativistic 
wave functions of $ B_c $ mesons \cite{apm5,apm6}:
\begin{equation}
\label{eq1}
{\cal M}(k_1,k_2,P,Q)=\int\frac{d{\bf p}}{(2\pi)^3}\Psi_{\cal V,P}(p,P)
\int\frac{d{\bf q}}{(2\pi)^3}\Psi_{\cal V,P}(q,Q){\cal T}(p,q,P,Q),
\end{equation}
where $k_1$, $k_2$ are four momenta of initial photons, $P$, $Q$ are four momenta
of final mesons. ${\cal T}(p,q,P,Q)$ is the production amplitude of four free quarks and
antiquarks in $\gamma\gamma$ interaction, $\Psi_{\cal V,P}(p,P)$ is the wave
function of $B_c$ meson. A superscript ${\cal P}$ indicates a pseudoscalar 
$B_c$ meson, a superscript ${\cal V}$ indicates a vector $B_c$ meson.
Four-momenta of the produced quarks and antiquarks can be expressed through
total and relative four-momenta in the form:
\begin{equation}
\label{eq2}
p_1=\eta_{1}P+p,~p_2=\eta_{2}P-p,~(p\cdot P)=0,~
\eta_{1,2}=\frac{M_{B_c}^2\pm m_1^2\mp m_2^2}{2M_{B_c}^2},
\end{equation}
\begin{displaymath}
q_1=\rho_{1}Q+q,~q_2=\rho_{2}Q-q,~(q\cdot Q)=0,~
\rho_{1,2}=\frac{M_{B_c}^2\pm m_1^2\mp m_2^2}{2M_{B_c}^2},
\end{displaymath}
where $M_{B_c}$ is the mass of pseudoscalar or vector $B_c^+$ ($B_c^{\ast +}$) meson.
Relative four-momenta of quarks $p=L_P(0,{\bf p})$ and
$q=L_P(0,{\bf q})$ are obtained from the rest frame four-momenta $(0,{\bf p})$ and $(0,{\bf q})$ by the Lorentz transformation to the system moving with the momenta $P$ and $Q$.
Relativistic coefficients $\eta_{1,2}$ and $\rho_{1,2}$ are taken in such a way that 
the following conditions  of orthogonality are satisfied: $(pP)=0$, $(qQ)=0$.
In the nonrelativistic approximation, when we neglect the binding energies of quarks 
in the meson, the coefficients $\rho_{1,2}\approx\eta_{1,2}\approx r_{1,2}=m_{1,2}/(m_1+m_2)$.
In what follows, we denote the mass of the b-quark $m_1$ and the mass of c-quark $m_2$.
In the case of $B_c$ mesons the values $r_1=0.76$, $r_2=0.24$ are very close to
$\eta_1=0.77$, $\rho_1=0.77$, $\eta_2=0.23$, $\rho_2=0.23$. 
Therefore, the use of coefficients $r_{1,2}$ is justified. It greatly simplifies 
the form of intermediate expressions for amplitudes.
It is useful to recall that
in the Bethe-Salpeter approach the initial production amplitude has a form of convolution 
of the truncated amplitude with two Bethe-Salpeter (BS) $B_c$ meson wave functions.
The presence of the $\delta (p\cdot P)$ function in this case allows us to make the integration 
over relative energy $p^0$. In the rest frame of a bound state the condition
$p^0=0$ allows to eliminate the relative energy from the BS wave function.

The amplitudes shown in Fig.~\ref{fig1} differ by replacing the lines of b-quarks and c-quarks, 
as well as by rearrangement of the initial photons.
We present analytical formulas for some amplitudes in the form 
(one representative from subclasses of amplitudes shown in Fig.~\ref{fig1}):
\begin{equation}
\label{eq3}
{\cal T}_1(p,q,P,Q)=16\pi^2\alpha_{s,c}\alpha Q_b^2\varepsilon^\mu_1(k_1)\varepsilon^\nu_2(k_2)
D^{ab}_{\lambda\sigma}(l_1)
\bar u_1^i(p_1)\gamma^\lambda T^a_{ij}\frac{\hat r-\hat q_1-m_1}{(r-q_1)^2-m_1^2}\gamma^\mu
\times
\end{equation}
\begin{displaymath}
\frac{\hat k_2-\hat q_1+m_1}{(k_2-q_1)^2-m_1^2}\gamma^\nu v_1^j(q_1)
\bar u_2^m(q_2)\gamma^\sigma T^b_{mn} v_2^n(p_2),
\end{displaymath}
\begin{equation}
\label{eq4}
{\cal T}_2(p,q,P,Q)=16\pi^2\alpha_{s,c}\alpha Q_b^2\varepsilon^\mu_1(k_1)\varepsilon^\nu_2(k_2)
D^{ab}_{\lambda\sigma}(l_1)
\bar u_1^i(p_1)\gamma^\mu T^a_{ij}\frac{\hat p_1-\hat k_1-m_1}{(p_1-k_1)^2-m_1^2}\gamma^\nu
\times
\end{equation}
\begin{displaymath}
\frac{\hat p_1-\hat r+m_1}{(p_1-r)^2-m_1^2}\gamma^\lambda v_1^j(q_1)
\bar u_2^m(q_2)\gamma^\sigma T^b_{mn} v_2^n(p_2),
\end{displaymath}
\begin{equation}
\label{eq5}
{\cal T}_3(p,q,P,Q)=16\pi^2\alpha_{s,c}\alpha Q_b^2\varepsilon^\mu_1(k_1)\varepsilon^\nu_2(k_2)
D^{ab}_{\lambda\sigma}(l_1)
\bar u_1^i(p_1)\gamma^\mu T^a_{ij}\frac{\hat p_1-\hat k_1+m_1}{(p_1-k_1)^2-m_1^2}\gamma^\lambda
\times
\end{equation}
\begin{displaymath}
\frac{\hat k_2-\hat q_1+m_1}{(k_2-q_1)^2-m_1^2}\gamma^\nu v_1^j(q_1)
\bar u_2^m(q_2)\gamma^\sigma T^b_{mn} v_2^n(p_2),
\end{displaymath}
\begin{equation}
\label{eq6}
{\cal T}_4(p,q,P,Q)=16\pi^2\sqrt{\alpha_{s,c}\alpha_{s,b}}
\alpha Q_c Q_b\varepsilon^\mu_1(k_1)\varepsilon^\nu_2(k_2)
D^{ab}_{\lambda\sigma}(l_2)
\bar u_1^i(p_1)\gamma^\mu T^a_{ij}\frac{\hat p_1-\hat k_1+m_1}{(p_1-k_1)^2-m_1^2}\gamma^\lambda
 v_1^j(q_1)\times
\end{equation}
\begin{displaymath}
\bar u_2^m(q_2)\gamma^\sigma T^b_{mn}
\frac{\hat k_2-\hat p_2+m_2}{(k_2-p_2)^2-m_2^2}\gamma^\nu v_2^n(p_2),
\end{displaymath}
\begin{equation}
\label{eq7}
{\cal T}_5(p,q,P,Q)=16\pi^2\sqrt{\alpha_{s,c}\alpha_{s,b}}
\alpha Q_c Q_b\varepsilon^\mu_1(k_1)\varepsilon^\nu_2(k_2)
D^{ab}_{\lambda\sigma}(l_2)
\bar u_1^i(p_1)\gamma^\mu T^a_{ij}\frac{\hat p_1-\hat k_1+m_1}{(p_1-k_1)^2-m_1^2}\gamma^\lambda
 v_1^j(q_1)\times
\end{equation}
\begin{displaymath}
\bar u_2^m(q_2)\gamma^\nu T^b_{mn}
\frac{\hat q_2-\hat k_2+m_2}{(q_2-k_2)^2-m_2^2}\gamma^\sigma v_2^n(p_2),
\end{displaymath}
where $l_1=r-p_1-p_1$, $l_2=k_2-p_2-q_2$. $\varepsilon_1(k_1)$ and $\varepsilon_2(k_2)$ 
are the polarization vectors of initial photons. 
$D^{ab}_{\lambda\sigma}(l)=\delta^{ab}D_{\lambda\sigma}(l)$ 
is the gluon propagator. $T^a$ is the SU(3) generator in the fundamental
representation.
The values of the strong coupling constant
at different energies are indicated $\alpha_{s,c}=\alpha_s(s\frac{m_2}{m_1+m_2})$ and
$\alpha_{s,b}=\alpha_s(s\frac{m_1}{m_1+m_2})$, $Q_c=2/3$ and $Q_b=-1/3$ are the charges
of heavy quarks,
$u^i_{1,2}$, $v^j_{1,2}$ are wave functions of free quarks and anti-quarks.
Color factor is equal to $\frac{\delta_{ni}}{\sqrt{3}}T^a_{ij}T^a_{jk}\frac{\delta_{kn}}
{\sqrt{3}}=\frac{4}{3}$ (the color part of the meson wave function is $\delta^{ik}/\sqrt{3}$).

Further transformation of the matrix elements in \eqref{eq1} is connected with the 
transformation law of the wave function of the bound state of quarks upon transition 
from the meson rest system to a moving reference frame with momenta Q and P. 
This law was obtained in the framework of the Bethe-Salpeter method in \cite{brodsky}, 
and in the framework of the three-dimensional quasipotential approach in \cite{faustov}. 
Since the quasipotential method is used in this work, the transformation 
of the wave function of mesons is represented as:
\begin{equation}
\label{eq8}
\Psi_{P}^{\rho\omega}({\bf p})=D_1^{1/2,~\rho\alpha}(R^W_{L_{P}})
D_2^{1/2,~\omega\beta}(R^W_{L_{P}})\Psi_{0}^{\alpha\beta}({\bf p}),
\end{equation}
\begin{displaymath}
\bar\Psi_{P}^{\lambda\sigma}({\bf p})
=\bar\Psi^{\varepsilon\tau}_{0}({\bf p})D_1^{+~1/2,~\varepsilon
\lambda}(R^W_{L_{P}})D_2^{+~1/2,~\tau\sigma}(R^W_{L_{P}}),
\end{displaymath}
where $R^W$ is the Wigner rotation, $L_{P}$ is the Lorentz boost
from the meson rest frame to a moving one. The rotation matrix $D^{1/2}(R)$ is determined by
\begin{equation}
\label{eq9}
{1 \ \ \,0\choose 0 \ \ \,1}D^{1/2}_{1,2}(R^W_{L_{P}})=
S^{-1}({\bf p}_{1,2})S({\bf P})S({\bf p}),
\end{equation}
where the Lorentz transformation matrix of the Dirac spinor is
\begin{equation}
\label{eq10}
S({\bf p})=\sqrt{\frac{\epsilon(p)+m}{2m}}\bigl(1+\frac{(\bm{\alpha}
{\bf p})} {\epsilon(p)+m}\bigr).
\end{equation}
Further transformations of the amplitudes \eqref{eq3}-\eqref{eq7} entering in \eqref{eq1}
should be carried out by means of the following expressions:
\begin{equation}
\label{eq11}
S_{\alpha\beta}(\Lambda)u^\lambda_\beta(p)=\sum_{\sigma=\pm 1/2}
u^{\sigma}_\alpha(\Lambda p)D^{1/2}_{\sigma\lambda}(R^W_{\Lambda p}),
\end{equation}
\begin{displaymath}
\bar u^\lambda_\beta(p)S^{-1}_{\beta\alpha}(\Lambda)=\sum_{\sigma=\pm 1/2}
D^{+~1/2}_{\lambda\sigma}(R^W_{\Lambda p})\bar u^\sigma_\alpha(\Lambda p).
\end{displaymath}
When constructing the production amplitudes of vector and pseudoscalar mesons, 
special projection operators on these states are used, which are constructed 
from the wave functions of quarks in the rest system. Therefore, the transformation 
formulas for Dirac bispinors of the following form are needed:
\begin{eqnarray}
\label{eq12}
\bar u_1({\bf p})=\bar u_1(0)\frac{(\hat
p'_1+m_1)}{\sqrt{2\epsilon_1({\bf p}) (\epsilon_1({\bf
p})+m_1)}},~~p'_1=(\epsilon_1,{\bf p}),\cr\cr v_2(-{\bf
p})=\frac{(\hat p'_2-m_2)}{\sqrt{2\epsilon_2({\bf
p})(\epsilon_2({\bf p})+ m_2)}}v_2(0),~~p'_2=(\epsilon_2,-{\bf p}).
\end{eqnarray}
After that we introduce the projection operators $\hat\Pi^{{\cal P},{\cal V}}$
on the heavy quark bound states $(c\bar b)$, $(b\bar c)$ with total spin 0 or 1:
\begin{equation}
\label{eq13}
\hat\Pi^{{\cal P},{\cal V}}=[v_2(0)\bar
u_1(0)]_{s=0,1}=\gamma_5(\hat\epsilon^\ast)\frac{1+\gamma^0}{2\sqrt{2}}.
\end{equation}

Total amplitude of pair $B_c$ meson production can be presented after such transformations 
as follows:
\begin{equation}
\label{eq14}
{\cal M}(k_1,k_2,P,Q)=\frac{16\pi^2\alpha M_{{\cal V,P}}}{3}
\int\frac{d{\bf p}}{(2\pi)^3}\int\frac{d{\bf q}}{(2\pi)^3}
Sp\left\{\Psi^{\cal V,P}_{B_c}(p,P)\Gamma^{(1)}\Psi^{\cal V,P}_{B_c}(q,Q)
\Gamma^{(2)}\right\},
\end{equation}
where $\Gamma^{(1,2)}$ are the vertex functions. For the amplitudes \eqref{eq3}-\eqref{eq7}
they have the form:
\begin{equation}
\label{eq15}
\Gamma^{(1)}_{ij}\Gamma^{(2)}_{kl}=\Bigl\{\alpha_{s,c} Q_b^2 D^{\lambda\sigma}(l_1)
\Bigl[\gamma^\lambda
\frac{\hat r-\hat q_1+m_1}{(r-q_1)^2-m_1^2}\gamma^\mu \frac{\hat k_2-\hat q_1+m_1}
{(k_2-q_1)^2-m_1^2}\gamma^\nu \Bigr]_{ij}\gamma^\sigma_{kl}+
\end{equation}
\begin{displaymath}
\alpha_{s,c} Q_b^2 D^{\lambda\sigma}(l_1)
\Bigl[\gamma^\mu
\frac{\hat p_1-\hat k_1+m_1}{(p_1-k_1)^2-m_1^2}\gamma^\nu \frac{\hat p_1-\hat r+m_1}
{(p_1-r)^2-m_1^2}\gamma^\lambda \Bigr]_{ij}\gamma^\sigma_{kl}+
\end{displaymath}
\begin{displaymath}
\alpha_{s,c} Q_b^2 D^{\lambda\sigma}(l_1)
\Bigl[\gamma^\mu
\frac{\hat p_1-\hat k_1+m_1}{(p_1-k_1)^2-m_1^2}\gamma^\lambda \frac{\hat k_2-\hat q_1+m_1}
{(k_2-q_1)^2-m_1^2}\gamma^\nu \Bigr]_{ij}\gamma^\sigma_{kl}+
\end{displaymath}
\begin{displaymath}
\sqrt{\alpha_{s,c}\alpha_{s,b}} Q_b Q_c D^{\lambda\sigma}(l_2)
\Bigl[\gamma^\mu
\frac{\hat p_1-\hat k_1+m_1}{(p_1-k_1)^2-m_1^2}\gamma^\lambda \Bigr]_{ij}
\Bigl[\gamma^\sigma \frac{\hat k_2-\hat p_2+m_2}
{(k_2-p_2)^2-m_2^2}\gamma^\nu \Bigr]_{kl}+
\end{displaymath}
\begin{displaymath}
\sqrt{\alpha_{s,c}\alpha_{s,b}} Q_b Q_c D^{\lambda\sigma}(l_2)
\Bigl[\gamma^\mu
\frac{\hat p_1-\hat k_1+m_1}{(p_1-k_1)^2-m_1^2}\gamma^\lambda \Bigr]_{ij}
\Bigl[\gamma^\nu \frac{\hat q_2-\hat k_2+m_2}
{(q_2-k_2)^2-m_2^2}\gamma^\sigma \Bigr]_{kl}
\Bigr\}\varepsilon^\mu_1(k_1)\varepsilon^\nu_2(k_2).
\end{displaymath}
The transition wave functions $\Psi^{\cal V}_{B^\ast_c}(q,Q)$ and 
$\Psi^{\cal P}_{B_c}(p,P)$ (form factors) have the following form:
\begin{eqnarray}
\label{eq16}
\Psi^{\cal P}_{B_c}(p,P)&=&\frac{\Psi^0_{B_c}({\bf p})}{
\sqrt{\frac{\epsilon_1(p)}{m_1}\frac{(\epsilon_1(p)+m_1)}{2m_1}
\frac{\epsilon_2(p)}{m_2}\frac{(\epsilon_2(p)+m_2)}{2m_2}}}
\left[\frac{\hat v_1-1}{2}+\hat
v_1\frac{{\bf p}^2}{2m_2(\epsilon_2(p)+ m_2)}-\frac{\hat{p}}{2m_2}\right]\cr
&&\times\gamma_5(1+\hat v_1) \left[\frac{\hat
v_1+1}{2}+\hat v_1\frac{{\bf p}^2}{2m_1(\epsilon_1(p)+
m_1)}+\frac{\hat{p}}{2m_1}\right],
\end{eqnarray}
\begin{eqnarray}
\label{eq17}
\Psi^{\cal V}_{B^\ast_c}(q,Q)&=&\frac{\Psi^0_{B^\ast_c}({\bf q})}
{\sqrt{\frac{\epsilon_1(q)}{m_1}\frac{(\epsilon_1(q)+m_1)}{2m_1}
\frac{\epsilon_2(q)}{m_2}\frac{(\epsilon_2(q)+m_2)}{2m_2}}}
\left[\frac{\hat v_2-1}{2}+\hat v_2\frac{{\bf q}^2}{2m_1(\epsilon_1(q)+
m_1)}+\frac{\hat{q}}{2m_1}\right]\cr &&\times\hat{\varepsilon}_{\cal
V}(Q,s_z)(1+\hat v_2) \left[\frac{\hat v_2+1}{2}+\hat
v_2\frac{{\bf q}^2}{2m_2(\epsilon_2(q)+ m_2)}-\frac{\hat{q}}{2m_2}\right],
\end{eqnarray}
where the four-vector $p_\mu$ convolution with the Dirac $\gamma^\mu$ matrix is indicated by
$\hat p$,
$v_1=P/M_{B_c}$, $v_2=Q/M_{B_c}$;
$\varepsilon_{\cal V}(Q,s_z)$ is the polarization vector of the $B^{\ast-}_c(1^-)$ meson,
relativistic quark energies $\epsilon_{1,2}(p)=\sqrt{p^2+m_{1,2}^2}$.
The matrix element \eqref{eq14} contains the integration over the quark relative momenta
${\bf p}$ and ${\bf q}$. The result of the integration in \eqref{eq14} is determined
by the bound state wave function but not the substitutions 
$M_{B_c}=\epsilon_1({\bf p})+\epsilon_2({\bf p})$ and
$M_{B^\ast_c}=\epsilon_1({\bf q})+\epsilon_2({\bf q})$. 
Expressions \eqref{eq16} and \eqref{eq17} include the meson wave functions in the rest frame
$\Psi^0_{B_c}({\bf p})$ and spin projection operators with relativistic factors 
of quarks. Another form of spin projector for $(c\bar c)$ system was used in \cite{bodwin2002} 
with heavy quark momenta lying on the mass shell.
The functions \eqref{eq16}-\eqref{eq17} describe a transition of
heavy quark-antiquark pair from free state to the bound state, so, they can be called 
the transition form factors. It is important to emphasize that a relativistic consideration 
of the production and decay of heavy quarkoniums is necessary to obtain the values 
of the observed quantities \cite{bodwin1,apm3,apm8,apm9}.

We have also the dependence on relative momenta $p$ and $q$ in vertex functions $\Gamma^{(1)}$,
$\Gamma^{(2)}$. The denominators of expression \eqref{eq15} can be simplified as follows:
\begin{equation}
\label{eq18}
(r-q_1)^2-m_1^2\approx r^2-2r_1rQ-2rq\approx r_2s^2,~~
(k_1-q_1)^2-m_1^2\approx -r_1s(P^0+|{\bf P}|\cos\theta),
\end{equation}
\begin{displaymath}
(k_2-q_1)^2-m_1^2\approx -2r_1k_2 Q=-2r_1 k_1P=-r_1s(P^0-|{\bf P}|\cos\theta),
\end{displaymath}
where $\theta$ is the angle between photon ${\bf k}_1$ and meson ${\bf P}$ three momenta.
so, we neglect here relative momenta p and q because corresponding corrections 
have the form $p^2/s^2$, $q^2/s^2$ and are very small in size at $s>2 M_{B_c}$.
We take into account exactly the corrections $p^2/m_{1,2}^2$, $q^2/m_{1,2}^2$ in
the numerator of relativistic amplitudes.

The appearance of a large number of the Feynman amplitudes describing the production 
of a pair of $ B_c $ mesons makes the use of the Form package \cite{form} especially important. 
The total contribution of 20 amplitudes is conveniently represented as a sum of 5 parts, 
since 4 of 20 amplitudes each have a similar mathematical structure.
The total amplitudes of pair production can be represented as a result in the form:
\begin{equation}
\label{eq19}
{\cal T}_{{\cal PP}}={\cal T}^{\mu\nu}_{{\cal PP}}\varepsilon_1^\mu(k_1)\varepsilon_2^\nu(k_2),~~~
{\cal T}^{\mu\nu}_{{\cal PP}}=\frac{A_1^{\mu\nu}}{r_1^3r_2^3s^5(P^0-|{\bf P}|z)}+
\frac{A_2^{\mu\nu}}{r_1^2r_2^2s^4(P^0-|{\bf P}|z)^2}+
\end{equation}
\begin{displaymath}
\frac{A_3^{\mu\nu}}{r_1^3r_2^3s^5(P^0+|{\bf P}|z)}+
\frac{A_4^{\mu\nu}}{r_1^2r_2^2s^4(P^0+|{\bf P}|z)^2}+
\frac{A_5^{\mu\nu}}{r_1^2r_2^2s^4(P^0-|{\bf P}|z)(P^0+|{\bf P}|z)},
\end{displaymath}
\begin{equation}
\label{eq21}
{\cal T}_{{\cal VV}}={\cal T}^{\mu\nu\epsilon_1\epsilon_2}_{{\cal VV}}\varepsilon_1^\mu(k_1)
\varepsilon_2^\nu(k_2)\varepsilon_{\cal V}^{\epsilon_1}(P)\varepsilon_{\cal V}^{\epsilon_2}(Q),
~~~{\cal T}^{\mu\nu\epsilon_1\epsilon_2}_{{\cal VV}}=
\frac{B_1^{\mu\nu\epsilon_1\epsilon_2}}{r_1^3r_2^3s^5(P^0-|{\bf P}|z)}+
\end{equation}
\begin{displaymath}
\frac{B_2^{\mu\nu\epsilon_1\epsilon_2}}{r_1^2r_2^2s^4(P^0-|{\bf P}|z)^2}+
\frac{B_3^{\mu\nu\epsilon_1\epsilon_2}}{r_1^3r_2^3s^5(P^0+|{\bf P}|z)}+
\frac{B_4^{\mu\nu\epsilon_1\epsilon_2}}{r_1^2r_2^2s^4(P^0+|{\bf P}|z)^2}+
\frac{B_5^{\mu\nu\epsilon_1\epsilon_2}}{r_1^2r_2^2s^4(P^0-|{\bf P}|z)(P^0+|{\bf P}|z)},
\end{displaymath}
where $\varepsilon_1^\mu(k_1)$, $\varepsilon_2^\nu(k_2)$, $\varepsilon_{\cal V}^{\epsilon_1}(P)$
are the polarization vectors of photons and vector meson, $z=\cos\theta$. 
We introduce the angle
$\theta$ between the photon momentum ${\bf k_1}$ and momentum ${\bf P}$ of $B_c$ meson.
The tensors appearing in \eqref{eq19}, \eqref{eq21} are very cumbersome when taking 
into account relativistic 
corrections. In the Appendix B, we write out the explicit form of these functions 
in the case of the production of a pair of pseudoscalar mesons.

To calculate the cross section we have to sum the squared modulus of the amplitude 
upon all polarizations using the following relations for final vector mesons 
and initial photons correspondingly ($v_1=P/M_{B_c}$, $v_2=Q/M_{B_c}$):
\begin{equation}
\label{eq22}
\sum_{\lambda}\varepsilon_{P}^{(\lambda)\mu}
{\varepsilon_{P}^\ast}^{(\lambda)\nu} = v_1^\mu v_1^\nu-g^{\mu\nu},
\sum_{\lambda}\varepsilon_{Q}^{(\lambda)\mu}
{\varepsilon_{Q}^\ast}^{(\lambda)\nu} = v_2^\mu v_2^\nu-g^{\mu\nu},
\sum_{\lambda}\varepsilon_{1,2}^{(\lambda)\mu} \varepsilon_{1,2}^{\ast\;(\lambda)\nu}=
\frac{k_1^\mu k_2^\nu+k_1^\nu k_2^\mu}{k_1\cdot k_2}-g^{\mu\nu}.
\end{equation}

After calculating the squared amplitude modulus, we obtain the differential 
effective cross section $d\sigma/d\cos\theta$ ($z=\cos\theta$)
\begin{equation}
\label{eq23}
d\sigma=\frac{1}{16\pi}|{\cal M}|^2\frac{|{\bf P}|}{s^3} dz
\end{equation}
as a function of center-of-mass energy $s$. It depends on a number of parameters 
including heavy quark masses and relativistic corrections.
$|{\bf P}|=\sqrt{(s^2-4M^2_{B_c})/4}$ 
is the meson three momentum in center-of-mass frame.
The technique of further transformations of $|{\cal M}|^2$ is described in detail 
in our previous works \cite{apm1,apm2}. 
Let us briefly repeat the basic elements of our transformations and the used quark model.
One of the very important parameters on which the numerical value of the production 
cross section depends (the cross section is proportional to the 4th power of this 
parameter) is the value of the bound state wave function at the origin. 
In the framework of the relativistic quark model, we have the following generalization of the nonrelativistic value $\Psi^{0,nr}_{B_c}(0)$:
\begin{equation}
\label{eq24}
\Psi^0_{B_c}(0)=\int \sqrt{\frac{(\epsilon_1(p)+m_1)(\epsilon_2(p)+m_2)}
{2\epsilon_1(p)\cdot 2\epsilon_2(p)}}\Psi^0_{B_c}({\bf p})\frac{d{\bf p}}{(2\pi)^3}.
\end{equation}
After the trace calculation in \eqref{eq14} we preserve in $|{\cal M}|^2$ only
relativistic corrections proportional to second degree of ${\bf p}$ and ${\bf q}$.
Next we express ${\bf p}^2$, ${\bf q}^2$ in powers of relativistic factors
$C_{nk}=[(m_1-\epsilon_1(p))/(m_1+\epsilon_1(p))]^n
[(m_2-\epsilon_2(q))/(m_2+\epsilon_2(q))]^k$ ( $n$, $k$ are integers and half-integers 
with $n+k\leq 1$):
\begin{equation}
\label{eq25}
\frac{{\bf p}^2}{4m_1^2}=\bigl(\frac{\epsilon_1-m_1}{\epsilon_1+m_1}\bigr)^{1/2}+
\bigl(\frac{\epsilon_1-m_1}{\epsilon_1+m_1}\bigr)^{3/2}+
\bigl(\frac{\epsilon_1-m_1}{\epsilon_1+m_1}\bigr)^{5/2}...,
\end{equation}
\begin{equation}
\label{eq26}
\frac{{\bf p}^2}{4m_2^2}=\bigl(\frac{\epsilon_2-m_2}{\epsilon_2+m_2}\bigr)^{1/2}+
\bigl(\frac{\epsilon_2-m_2}{\epsilon_2+m_2}\bigr)^{3/2}+
\bigl(\frac{\epsilon_2-m_2}{\epsilon_2+m_2}\bigr)^{5/2}...,
\end{equation}
\begin{equation}
\label{eq27}
\frac{{\bf p}^2}{4m_1m_2}=\bigl(\frac{(\epsilon_1-m_1)(\epsilon_2-m_2)}
{(\epsilon_1+m_1)(\epsilon_2+m_2)}\bigr)^{1/2}
[1+\frac{(\epsilon_1-m_1)}{(\epsilon_1+m_1)}+\frac{(\epsilon_2-m_2)}{(\epsilon_2+m_2)}+...].
\end{equation}
Last equation \eqref{eq27} is used to maintain the symmetry over quarks 1 and 2
in some terms.
Finally, we introduce special relativistic parameters $\omega_{nk}$ which are defined as follows
in terms of momentum integrals $I^{nk}$ and calculated in the quark model:
\begin{equation}
\label{eq28}
I^{nk}_{B_c}=\int_0^\infty q^2R_{B_c}(q)\sqrt{\frac{(\epsilon_1(q)+m_1)(\epsilon_2(q)+m_2)}
{2\epsilon_1(q)\cdot 2\epsilon_2(q)}}
\bigl(\frac{\epsilon_1(q)-m_1}{\epsilon_1(q)+m_1}\bigr)^n
\bigl(\frac{\epsilon_2(q)-m_2}{\epsilon_2(q)+m_2}\bigr)^k dq,
\end{equation}
\begin{equation}
\label{eq29}
\omega^{B_c}_{10}=\frac{I_{B_c}^{10}}{I_{B_c}^{00}},~~~
\omega^{B_c}_{01}=\frac{I_{B_c}^{01}}{I_{B_c}},~~~
\omega^{B_c}_{\frac{1}{2}\frac{1}{2}}=\frac{I_{B_c}^{\frac{1}{2}\frac{1}{2}}}
{I_{B_c}^{00}},
\end{equation}
where $R_{B_c}(q)$ is the radial wave function of the $B_c$ mesons 
in momentum space. With a good degree of accuracy, 
the main contribution to expansions \eqref{eq25}-\eqref{eq27} is made by the first terms, 
which we retain when calculating the cross sections.

\begin{figure}[t!]
\centering
\includegraphics[scale=0.8]{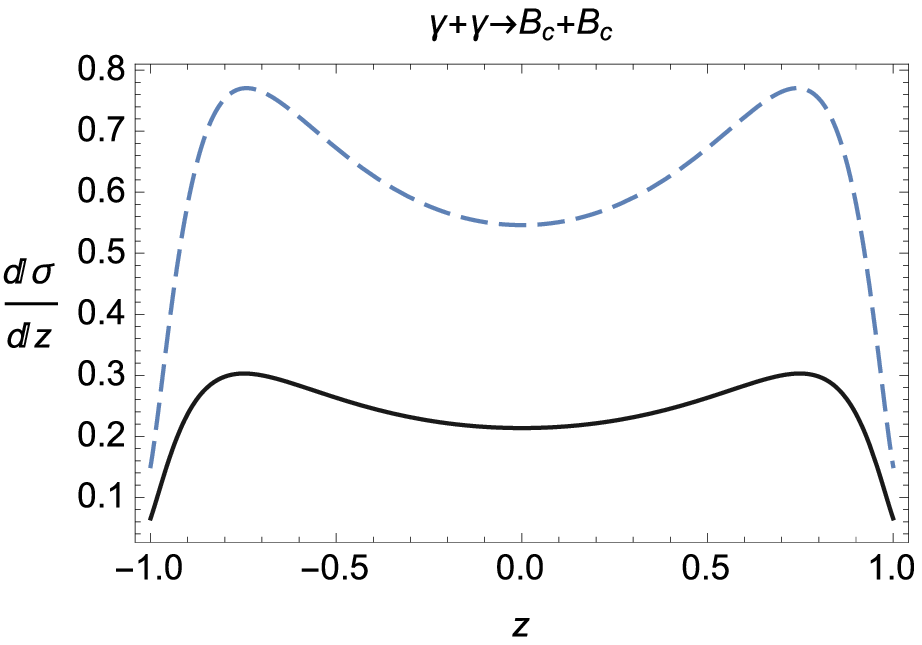}
\includegraphics[scale=0.8]{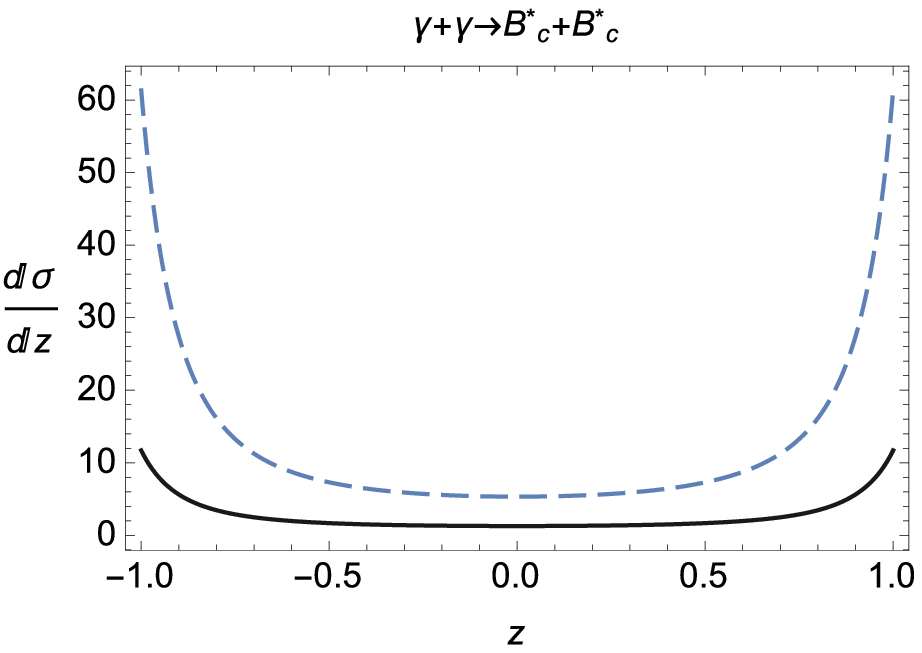}
\caption{The differential cross sections $d\sigma/dz$ ($z=\cos\theta$) in fb of pair 
$B_c$ meson production in $\gamma\gamma$ interaction at the center-of-mass energy
$s=20$ GeV (solid line). The dashed line shows nonrelativistic result.}
\label{fig2}
\end{figure}

Up to this point we have discussed the first group of relativistic corrections
to the pair $B_c$ meson production amplitude \eqref{eq14}. As already noted
they are connected with terms in \eqref{eq14} containing momenta of relative motion
${\bf p}$ and ${\bf q}$. But there exists another group of relativistic corrections
which appear in the interaction operator of heavy quarks b and c and, as a result,
in the bound state wave function $\Psi^0_{B_c}({\bf p})$ in \eqref{eq16}-\eqref{eq17}
\cite{rqm1,pot,repko2,repko3,rqm3,godfrey,lucha}.
For their calculation we use nonrelativistic Schr\"odinger equation in which the
particle interaction operator
\begin{equation}
\label{eq30}
H=H_0+\Delta U_1+\Delta U_2,~~~H_0=\sqrt{{\bf
p}^2+m_1^2}+\sqrt{{\bf p}^2+m_2^2}-\frac{4\tilde\alpha_s}{3r}+(Ar+B),
\end{equation}
consists of perturbative part $\Delta U_1(r)$ and $\Delta U_2(r)$:
\begin{equation}
\label{eq31}
\Delta U_1(r)=-\frac{\alpha_s^2}{3\pi r}\left[2\beta_0\ln(\mu
r)+a_1+2\gamma_E\beta_0
\right],~~a_1=\frac{31}{3}-\frac{10}{9}n_f,~~\beta_0=11-\frac{2}{3}n_f,
\end{equation}
\begin{equation}
\label{eq32}
\Delta U_2(r)=-\frac{2\alpha_s}{3m_1m_2r}\left[{\bf p}^2+\frac{{\bf
r}({\bf r}{\bf p}){\bf p}}{r^2}\right]+\frac{2\pi
\alpha_s}{3}\bigl(\frac{1}{m_1^2}+\frac{1}{m_2^2}\bigr)\delta({\bf r})+
\frac{4\alpha_s}{3r^3}\bigl(\frac{1}{2m_1^2}+\frac{1}{m_1m_2}\bigr)({\bf s}_1{\bf L})+
\end{equation}
\begin{displaymath}
+\frac{4\alpha_s}{3r^3}\bigl(\frac{1}{2m_2^2}+\frac{1}{m_1m_2}\bigr)({\bf s}_2{\bf L})
+\frac{32\pi\alpha_s}{9m_1m_2}({\bf s}_1{\bf s}_2)\delta({\bf r})+
\frac{4\alpha_s}{m_1m_2r^3}\left[\frac{({\bf s}_1{\bf r})({\bf s}_2{\bf r})}{r^2}-
\frac{1}{3}({\bf s}_1{\bf s}_2)\right]-
\end{displaymath}
\begin{displaymath}
-\frac{\alpha_s^2(m_1+m_2)}{m_1m_2r^2}\left[1-\frac{4m_1m_2}{9(m_1+m_2)^2}\right],
\end{displaymath}
and nonperturbative interaction $Ar+B+\Delta V^{hfs}_{conf}(r)$:
\begin{equation}
\label{eq33}
\Delta V^{hfs}_{conf}(r)=
f_V\frac{A}{8r}\left\{\frac{1}{m_1^2}+\frac{1}{m_2^2}+\frac{16}{3m_1m_2}({\bf s}_1{\bf s}_2)+
\frac{4}{3m_1m_2}\left[3({\bf s}_1 {\bf r}) ({\bf s}_2 {\bf r})-({\bf s}_1 {\bf s}_2)\right]\right\},
\end{equation}
where an additional parameter is taken equal to $f_V=0.9$,
${\bf L}=[{\bf r}\times{\bf p}]$, ${\bf s}_1$, ${\bf s}_2$ are spins of heavy quarks,
$n_f$ is the number of flavors, the Euler constant $\gamma_E\approx 0.577216$.
All parameters of the quark interaction operator \eqref{eq30}-\eqref{eq33}
were fixed as in previous papers \cite{apm1,apm2}.

\begin{table}[h]
\caption{Relativistic parameters entering in the production cross sections.}
\bigskip
\label{tb1}
\begin{ruledtabular}
\begin{tabular}{|c|c|c|c|c|c|c|}
$B_c$ &$n^{2S+1}L_J$ &$M_{B_c}$, GeV&$\Psi^0_{B_c}(0)$, GeV$^{3/2}$ & $\omega_{10}$ &$\omega_{01}$ &
$\omega_{\frac{1}{2}\frac{1}{2}}$   \\
meson  &     &    &    &    &     &      \\    \hline
$B_c$&$1^1S_0$ & 6.276 & 0.250 & 0.0489 & 0.0060  & 0.0171     \\  \hline
$B^\ast_c$  & $1^3S_1$  & 6.317 & 0.211 & 0.0540  & 0.0066   &  0.0188   \\  \hline
\end{tabular}
\end{ruledtabular}
\end{table}

Solving the Schr\"odinger equation numerically with the account terms 
\eqref{eq31}-\eqref{eq33}, 
we obtain the wave function of the bound state $ (b \bar c) $ with spin S = 0 and 1. 
Using this solution, numerical values ​​of parameters \eqref{eq24} and \eqref{eq29} are calculated 
and presented in the Table~\ref{tb1}. The plots in Fig.~\ref{fig2} show the type 
of differential effective production cross 
sections for a pair of $B_c$ mesons without and taking into account relativistic 
corrections. A significant decrease in the production cross sections taking into account 
relativistic corrections with respect to nonrelativistic cross sections is explained 
primarily by the decrease in the relativistic case in our model of the key parameter 
$ \Psi^0_{B_c}(0)$. The ratio 
$|\Psi^{0,nr}_{B_c}(0)|^4/||\Psi^{0}_{B_c}(0)|^4$
for the states with spin 0 and 1, 
which is 2.4 and 4.8, respectively, shows how many times the nonrelativistic cross 
section decreases due to this factor.

Table~\ref{tb1} shows the masses of $ B_c $ mesons (the states $^1S_0 $, $^3S_1 $) 
obtained in our model. They are in good agreement with both the experimental result 
$M(1^1S_0)=6.2749$ GeV \cite{pdg}
and theoretical calculations $M(1^3S_1)=6.332$ GeV \cite{rqm1}. 
It should be emphasized the multidirectional effect of various relativistic corrections. 
Those relativistic corrections ${\bf p}^2/m_{1,2}^2$, ${\bf q}^2/m_{1,2}^2$
that are taken into account in the production amplitudes 
\eqref{eq19}-\eqref{eq21} give an increase in the production cross sections from 20 to 
50\% for different mesons. But relativistic corrections in the interaction potential 
of heavy quarks lead to a significant decrease in $ \Psi^0_{B_c}(0) $ and, accordingly, 
to a decrease in the production cross sections. In general, this last factor is decisive.

\begin{figure}[t!]
\centering
\includegraphics[scale=0.75]{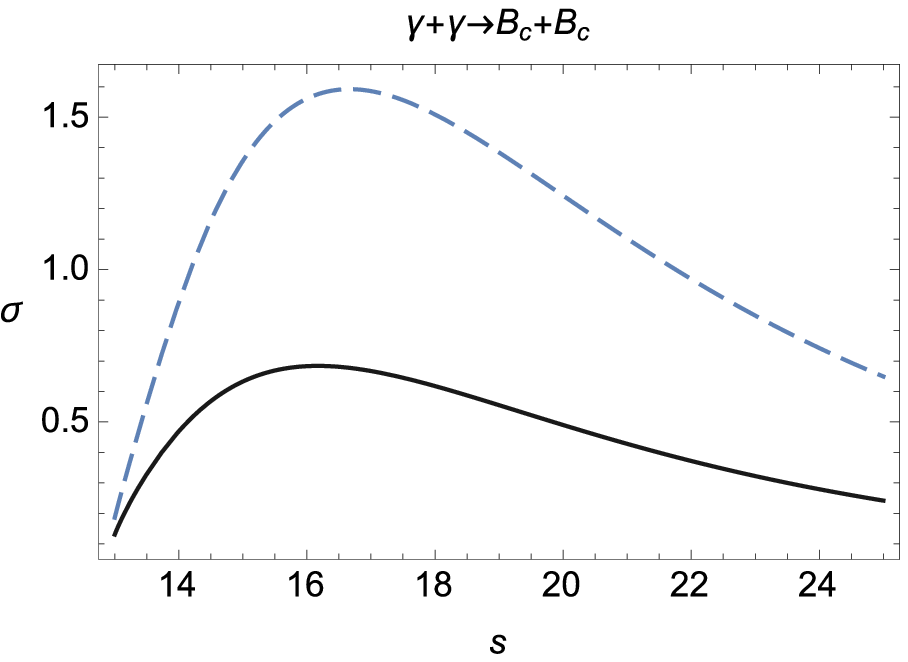}
\includegraphics[scale=0.75]{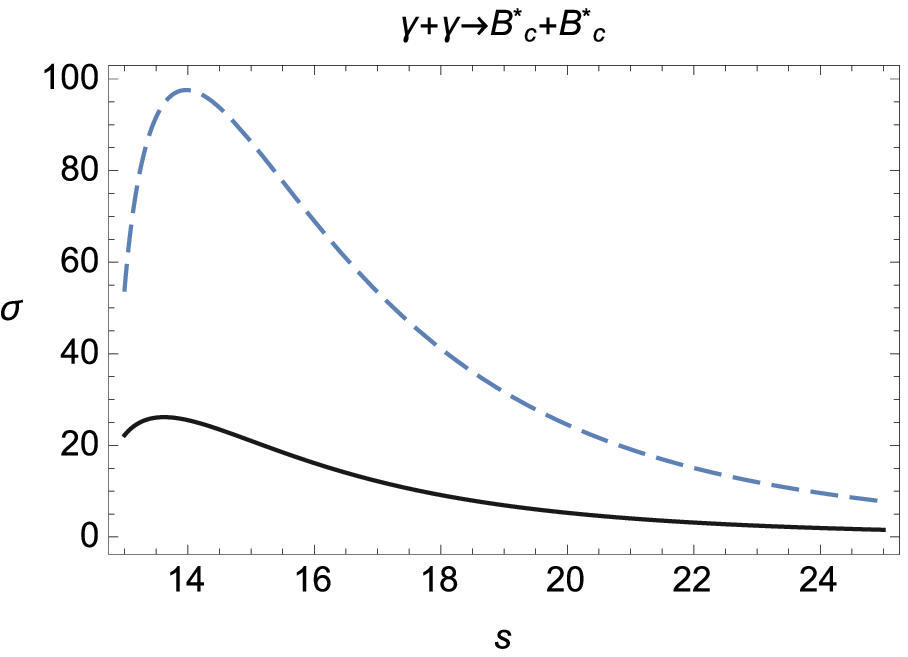}
\caption{The cross section $\sigma$ in fb of pair 
$B_c$ meson production in $\gamma\gamma$ interaction as a function of the center-of-mass energy
$s$ in GeV (solid line). The dashed line shows nonrelativistic result.}
\label{fig3}
\end{figure}

\section{Numerical results and conclusion}

While charmonium states have already been studied in sufficient detail, 
new excited states are being discovered, experimental data on the bound 
states $ (c \bar b) $, $ (\bar c b) $ are rather small. Therefore, the search for 
various physical processes for their observation is an important task.
The creation of a photonic collider based on ILC is of undoubted interest. 
The use of beams of real high-energy photons will allow us to study new 
processes of particle interaction and to verify existing theoretical models 
to describe them. One of these discussed processes is connected 
with the production of the Higgs boson and the study of various channels for its decay. 
Another promising direction of the $ \gamma \gamma $ interaction, which is studied 
in this work, is related with the production of bound states of heavy quarks.
The process of exclusive production of a pair of doubly heavy mesons or baryons is an important 
source of information on the interaction of heavy quarks. It can be used both 
to verify existing models for describing the bound states of particles, and to refine 
the values ​​of the fundamental parameters of the theory. All particles generated 
in this reaction can be reliably identified by their decay products.
Our relativistic approach to the study of quarkonium pair production reactions, 
formulated in \cite{apm1,apm2,apm3,apm4}, is extended in this work
to a new process $\gamma+\gamma\to B_c^++B_c^-$.

We obtain leading order relativistic amplitudes \eqref{eq1}, \eqref{eq19}, \eqref{eq21}
for the exclusive pair production of pseudoscalar and vector $B_c$ mesons
using the perturbative QCD and relativistic quark model.
Taking into account the smallness of relative momenta of heavy quarks we make an expansion of
the amplitude on p and q and hold the terms of the second order. 
There are two types of relativistic corrections related to the expansion in small parameters
${\bf p}^2/m_{1,2}^2$, ${\bf q}^2/m_{1,2}^2$ and ${\bf p}^2/s^2$, ${\bf q}^2/s^2$.
The second type of corrections occurs when the denominators of the heavy quark propagators 
are expanded at small relative momenta p, q. We neglect the corrections of the form 
${\bf p}^2/s^2$, ${\bf q}^2/s^2$
since at energies $ s> 15 $ GeV their numerical value is an order of magnitude smaller 
than the first. This somewhat simplifies the final form of the production cross sections.
It is also useful to note that the quark model used by us has certain universality, 
since the quark interaction operator contains the perturbative part based 
on calculations in the framework of the nonrelativistic QCD and the widely used 
nonperturbative part \cite{brambilla2011,rqm1,pot,repko2,repko3,godfrey,lucha}.

Performing analytical integration of the differential effective cross sections \eqref{eq23}
over the angle $\theta$, we obtain the total cross sections for the production of a pair 
of mesons. The corresponding expressions are very bulky, so we present here only 
nonrelativistic cross sections in the form:
\begin{equation}
\label{eq34}
\sigma_{{\cal PP}}=\frac{512\pi^3\alpha_{s,c}^2 Q_c^4\alpha^2
|{\bf P}|}{27r_1^6r_2^6\tilde s^{14}M^9}|\Psi^0_{{\cal P}}(0)|^4
\Biggl[\tilde s h_1(\tilde s)+\frac{1}{\sqrt{\tilde s^2-4}}\ln\frac{\tilde s-\sqrt{\tilde s^2-4}}
{\tilde s+\sqrt{\tilde s^2-4}}h_2(\tilde s)\Biggr],
\end{equation}
\begin{equation}
\label{eq35}
h_1(\tilde s)=
192 \bigl((q_{bc}-1)^2 \bigl(q_{bc}^2-34 q_{bc}+113\bigr) r_1^6+\bigl(36 q_{bc}^2-220 
q_{bc}+225\bigr) r_1^4-2
\bigl(q_{bc}^2-40 q_{bc}+80\bigr) r_1^3+
\end{equation}
\begin{displaymath}
2 \bigl(4 q_{bc}^3-63 q_{bc}^2+160 q_{bc}-101\bigr) r_1^5+5
(q_{bc}-1)^4 r_1^8-4 (q_{bc}-9) (q_{bc}-1)^3 r_1^7+(71-12 q_{bc}) r_1^2-18 r_1+2\bigr)-
\end{displaymath}
\begin{displaymath}
(r_1-1)^2r_1^2 \tilde s^6 \bigl(q_{bc}^4 \bigl(72 r_1^4-80 r_1^3-20 r_1^2+44 r_1-11\bigr)-4 q_{bc}^3 \bigl(72
r_1^4-112 r_1^3+36 r_1^2+24 r_1-11\bigr)+
\end{displaymath}
\begin{displaymath}
q_{bc}^2 \bigl(432 r_1^4-864 r_1^3+520 r_1^2-88
r_1-26\bigr)-4 q_{bc} \bigl(72 r_1^4-176 r_1^3+132 r_1^2-48 r_1+9\bigr)+72 r_1^4-208 r_1^3+
\end{displaymath}
\begin{displaymath}
172r_1^2-52 r_1+5\bigr)+16 (r_1-1) r_1\tilde s^2 \bigl(q_{bc}^4 r_1^3 \bigl(34 r_1^3-46 r_1^2+15
r_1-3\bigr)-2 q_{bc}^3 r_1^2 \bigl(68 r_1^4-148 r_1^3+98 r_1^2-
\end{displaymath}
\begin{displaymath}
27 r_1+6\bigr)+3 q_{bc}^2 \bigl(68
r_1^6-204 r_1^5+214 r_1^4-88 r_1^3+17 r_1^2-7 r_1+1\bigr)-2 q_{bc} (r_1-1)^2 \bigl(68
r_1^4-124 r_1^3+62 r_1^2+
\end{displaymath}
\begin{displaymath}
3 r_1-3\bigr)+(r_1-1)^3 r_1 \bigl(34 r_1^2-56 r_1+25\bigr)\bigr)-2 s^4
\bigl(q_{bc}^4 r_1^2 \bigl(72 r_1^6-64 r_1^5-144 r_1^4+228 r_1^3-113 r_1^2+18 r_1-3\bigr)
\end{displaymath}
\begin{displaymath}
-4q_{bc}^3 r_1^2 \bigl(72 r_1^6-176 r_1^5+160 r_1^4-46 r_1^3-23 r_1^2+16 r_1-3\bigr)+2 q_{bc}^2
r_1 \bigl(216 r_1^7-864 r_1^6+1568 r_1^5-1680 r_1^4+
\end{displaymath}
\begin{displaymath}
1175 r_1^3-558 r_1^2+173 r_1-30\bigr)-4
q_{bc} (r_1-1)^2 r_1 \bigl(72 r_1^5-256 r_1^4+360 r_1^3-274 r_1^2+119 r_1-24\bigr)+(r_1-1)^2
\end{displaymath}
\begin{displaymath}
\bigl(72 r_1^6-368 r_1^5+616 r_1^4-452 r_1^3+147 r_1^2-12 r_1-6\bigr)\bigr)+(q_{bc}-1)^4 (r_1-1)^2
r_1^2 \bigl(8 r_1^4-16 r_1^3+12 r_1^2-4 r_1+1\bigr)\tilde s^8,
\end{displaymath}
\begin{equation}
\label{eq36}
h_2(\tilde s)=
6 \bigl((q_{bc}-1)^2 (r_1-1)^2 r_1^2 s^8 \bigl(q_{bc}^2 \bigl(8 r_1^4-8 r_1^3+4 r_1-1\bigr)-2 q_{bc}
\bigl(8 r_1^4-16 r_1^3+8 r_1^2-1\bigr)+8 r_1^4-24 r_1^3+
\end{equation}
\begin{displaymath}
24 r_1^2-12 r_1+3\bigr)-2 (q_{bc}-1)^2
   (r_1-1) r_1\tilde s^6 \bigl(-16 \bigl(2 q_{bc}^2-3 q_{bc}+1\bigr) r_1^5+8 \bigl(4 q_{bc}^2-6 q_{bc}-1\bigr)
   r_1^4-
\end{displaymath}
\begin{displaymath}
16 \bigl(q_{bc}^2-q_{bc}-3\bigr) r_1^3+\bigl(5 q_{bc}^2-2 q_{bc}-51\bigr) r_1^2+\bigl(-q_{bc}^2+2
   q_{bc}+23\bigr) r_1+8 (q_{bc}-1)^2 r_1^6-4\bigr)+64 \bigl((q_{bc}-1)^2 
\end{displaymath}
\begin{displaymath}
\bigl(q_{bc}^2-34 q_{bc}+113\bigr)
r_1^6+\bigl(36 q_{bc}^2-220 q_{bc}+225\bigr) r_1^4-2 \bigl(q_{bc}^2-40 q_{bc}+80\bigr) r_1^3+2 \bigl(4q_{bc}^3-63 q_{bc}^2+160 q_{bc}-
\end{displaymath}
\begin{displaymath}
101\bigr) r_1^5+5 (q_{bc}-1)^4 r_1^8-4 (q_{bc}-9) (q_{bc}-1)^3
r_1^7+(71-12 q_{bc}) r_1^2-18 r_1+2\bigr)-2\tilde s^4 \bigl(4 (q_{bc}-1)^2 \bigl(5 q_{bc}^2-
\end{displaymath}
\begin{displaymath}
106 q_{bc}+173\bigr)
r_1^6+4 \bigl(3 q_{bc}^4+56 q_{bc}^3-330 q_{bc}^2+528 q_{bc}-257\bigr) r_1^5+\bigl(-15 q_{bc}^4-36
   q_{bc}^3+606 q_{bc}^2-1396 q_{bc}+
\end{displaymath}
\begin{displaymath}905\bigr) r_1^4+2 \bigl(3 q_{bc}^4-8 q_{bc}^3-66 q_{bc}^2+240
q_{bc}-233\bigr) r_1^3+\bigl(-q_{bc}^4+4 q_{bc}^3+6 q_{bc}^2-52 q_{bc}+123\bigr) 
r_1^2+40 (q_{bc}-1)^4r_1^8
\end{displaymath}
\begin{displaymath} 
-64 (q_{bc}-4) (q_{bc}-1)^3 r_1^7-8 (q_{bc}+1) r_1-2\bigr)+16\tilde s^2 \bigl((q_{bc}-1)^2 \bigl(15
q_{bc}^2-110 q_{bc}+71\bigr) r_1^6+\bigl(8 q_{bc}^2+16 q_{bc}-
\end{displaymath}
\begin{displaymath}109\bigr) r_1^2-\bigl(q_{bc}^2+2
q_{bc}-40\bigr) r_1+4 \bigl(q_{bc}^3-8 q_{bc}^2-6 q_{bc}+37\bigr) r_1^3+\bigl(-6 q_{bc}^4+78 q_{bc}^3-246
q_{bc}^2+202 q_{bc}-28\bigr) r_1^5+
\end{displaymath}
\begin{displaymath}
\bigl(q_{bc}^4-22 q_{bc}^3+109 q_{bc}^2-
52 q_{bc}-84\bigr) r_1^4+8(q_{bc}-1)^4 r_1^8-8 (q_{bc}-1)^3 (3 q_{bc}-5) r_1^7-6\bigr)\bigr),
\end{displaymath}
\begin{equation}
\label{eq40}
\sigma_{{\cal VV}}=\frac{1024\pi^3\alpha_{s,c}^2 Q_c^4\alpha^2
|{\bf P}|}{27r_1^6r_2^6\tilde s^{14}M^9}|\Psi^0_{{\cal V}}(0)|^4
\Biggl[\tilde s h_5(\tilde s)+\frac{1}{\sqrt{\tilde s^2-4}}\ln\frac{\tilde s-\sqrt{\tilde s^2-4}}
{\tilde s+\sqrt{\tilde s^2-4}}h_6(\tilde s)\Biggr],
\end{equation}
\begin{equation}
\label{eq41}
h_5(\tilde s)=
2 \bigl(q_{bc}^4+6 q_{bc}^2+1\bigr) (r_1-1)^2 r_1^2\tilde s^8+48 \bigl((q_{bc}-1)^2 
\bigl(17 q_{bc}^2-302q_{bc}+829\bigr) r_1^6+5 \bigl(54 q_{bc}^2-304 q_{bc}+
\end{equation}
\begin{displaymath}
303\bigr) r_1^4-10 \bigl(q_{bc}^2-52 q_{bc}+104\bigr)
r_1^3+2 \bigl(44 q_{bc}^3-495 q_{bc}^2+1160 q_{bc}-709\bigr) r_1^5+40 (q_{bc}-1)^4 r_1^8-4 (q_{bc}-1)^3 
\end{displaymath}
\begin{displaymath}
(11q_{bc}-69) r_1^7+(451-72 q_{bc}) r_1^2-114 r_1+13\bigr)+4 (r_1-1) r_1\tilde s^6 
\bigl(q_{bc}^4 r_1\bigl(r_1^3-4 r_1^2+10 r_1-7\bigr)+
\end{displaymath}
\begin{displaymath}
q_{bc}^3 r_1 \bigl(-4 r_1^3+12 r_1^2-17 r_1+15\bigr)+2
q_{bc}^2 r_1 \bigl(3 r_1^3-6 r_1^2+14 r_1-11\bigr)-q_{bc} \bigl(4 r_1^4-4 r_1^3+5
r_1^2+r_1-6\bigr)+
\end{displaymath}
\begin{displaymath}
r_1 \bigl(r_1^3+4 r_1-5\bigr)\bigr)+\tilde s^4 \bigl(8 (q_{bc}-1)^2 \bigl(9 q_{bc}^2-67
q_{bc}+121\bigr) r_1^6+12 \bigl(8 q_{bc}^2-7 q_{bc}-3\bigr) r_1-4 \bigl(16 q_{bc}^4+9 q_{bc}^3+
\end{displaymath}
\begin{displaymath}
240q_{bc}^2-581 q_{bc}+316\bigr) r_1^5+\bigl(83 q_{bc}^4+476 q_{bc}^3-358 q_{bc}^2-944 q_{bc}+843\bigr) r_1^4-4\bigl(6 q_{bc}^4+86 q_{bc}^3-131 q_{bc}^2+36 q_{bc}
\end{displaymath}
\begin{displaymath}
+53\bigr) r_1^3+\bigl(24 q_{bc}^4+72 q_{bc}^3-262
q_{bc}^2+296 q_{bc}-6\bigr) r_1^2+64 (q_{bc}-1)^4 r_1^8-128 (q_{bc}-3) 
(q_{bc}-1)^3 r_1^7+27\bigr)+
\end{displaymath}
\begin{displaymath}
4\tilde s^2\bigl(-4 (q_{bc}-1)^2 \bigl(25 q_{bc}^2-445 q_{bc}+963\bigr) r_1^6+2 \bigl(69 q_{bc}^2+162 q_{bc}-599\bigr)r_1^2-6 \bigl(4 q_{bc}^2+q_{bc}-37\bigr) r_1+
\end{displaymath}
\begin{displaymath}
4 \bigl(24 q_{bc}^3-51 q_{bc}^2-553 q_{bc}+852\bigr) r_1^3-2
\bigl(9 q_{bc}^4+209 q_{bc}^3-2532 q_{bc}^2+5339 q_{bc}-3025\bigr) r_1^5+
\bigl(15 q_{bc}^4-234 q_{bc}^3-
\end{displaymath}
\begin{displaymath}
1206q_{bc}^2+6656 q_{bc}-5775\bigr) r_1^4-208 (q_{bc}-1)^4 r_1^8+8 (q_{bc}-1)^3 
(37 q_{bc}-171) r_1^7-15\bigr),
\end{displaymath}
\begin{equation}
\label{eq42}
h_6(\tilde s)=
6 \bigl(16 \bigl((q_{bc}-1)^2 \bigl(17 q_{bc}^2-302 q_{bc}+829\bigr) r_1^6+5 \bigl(54 q_{bc}^2-304 q_{bc}+303\bigr)r_1^4-10 \bigl(q_{bc}^2-52 q_{bc}+
\end{equation}
\begin{displaymath}
104\bigr) r_1^3+2 \bigl(44 q_{bc}^3-495 q_{bc}^2+1160 q_{bc}-709\bigr)
r_1^5+40 (q_{bc}-1)^4 r_1^8-4 (q_{bc}-1)^3 (11 q_{bc}-69) r_1^7+(451-
\end{displaymath}
\begin{displaymath}
72 q_{bc}) r_1^2-114
r_1+13\bigr)+\tilde s^4 \bigl(8 (q_{bc}-1)^2 \bigl(7 q_{bc}^2-64 q_{bc}+147\bigr) 
r_1^6-8 \bigl(2 q_{bc}^2-8
q_{bc}+17\bigr) r_1+4 \bigl(7 q_{bc}^4+6 q_{bc}^3-
\end{displaymath}
\begin{displaymath}
348 q_{bc}^2+818 q_{bc}-483\bigr) r_1^5+\bigl(-43
q_{bc}^4+248 q_{bc}^3+294 q_{bc}^2-2272 q_{bc}+2057\bigr) r_1^4-4 \bigl(2 q_{bc}^4+30 q_{bc}^3+15 q_{bc}^2-
\end{displaymath}
\begin{displaymath}
266q_{bc}+361\bigr) r_1^3+\bigl(8 q_{bc}^4+24 q_{bc}^3+70 q_{bc}^2-352 q_{bc}+622\bigr) r_1^2+
64 (q_{bc}-1)^4
r_1^8-32 (q_{bc}-1)^3 (3 q_{bc}-13) r_1^7+9\bigr)
\end{displaymath}
\begin{displaymath}
-4\tilde s^2 \bigl(2 (q_{bc}-1)^2 \bigl(35 q_{bc}^2-372
q_{bc}+931\bigr) r_1^6-2 \bigl(23 q_{bc}^2+78 q_{bc}-540\bigr) r_1^2+2 \bigl(4 q_{bc}^2+q_{bc}-151\bigr)r_1-4 \bigl(8 q_{bc}^3-
\end{displaymath}
\begin{displaymath}
28 q_{bc}^2-271 q_{bc}+584\bigr) r_1^3+6 \bigl(q_{bc}^4+33 q_{bc}^3-366 q_{bc}^2+851
q_{bc}-519\bigr) r_1^5+\bigl(-5 q_{bc}^4+78 q_{bc}^3+430 q_{bc}^2-3232 q_{bc}+
\end{displaymath}
\begin{displaymath}
3315\bigr) r_1^4+96
(q_{bc}-1)^4 r_1^8-128 (q_{bc}-5) (q_{bc}-1)^3 r_1^7+39\bigr)+8 (q_{bc}-1)^2 
q_{bc} (r_1-1)^2 r_1^2\tilde  s^6\bigr),
\end{displaymath}
where $\tilde s=s/M=s/(m_1+m_2)$, $q_{bc}=Q_b\alpha_{s,c}/Q_c\alpha_{s,b}$.
From these expressions, the general structure of the production cross sections 
is clearly visible: they are determined by functions that are an expansion 
in two parameters $r_1$ and $q_{bc}$ at a given energy s.
The terms with relativistic corrections have the same form.

The type of total production cross sections $\sigma_{{\cal PP}}$ and $\sigma_{{\cal VV}}$ 
depending on center-of-mass energy s is shown 
in Fig.~\ref{fig3}, on which we demonstrate a comparison of results of nonrelativistic and 
relativistic calculation. 
In general, the type of cross sections for pair production of mesons remains 
the same as for other reactions $e^+e^-$ annihilation and pp-interaction. At the threshold, 
the cross sections vanish due to the $|{\bf P}|$ factor. In our calculations, we did not 
take into account the Coulomb interaction of particles in the final state.

The production of a pair of $B_c$ mesons in $e^+e^-$ annihilation was 
considered by us earlier in \cite{apm1,apm2}. In these works, both single-photon and 
two-photon meson pair production mechanisms were investigated. The cross section 
for the production of a pair of $B_c$ mesons in $ e^+ e^- $ annihilation in 
processes with two virtual photons is suppressed with respect to the cross section 
obtained in the single-photon mechanism by an additional $\alpha^2$ factor. In the reaction 
$ \gamma \gamma \to B_c^+ B_c^-$ with two real photons, the general factor 
in the cross sections \eqref{eq33}, \eqref{eq36} is the 
same as in electron-positron annihilation with one virtual photon. Nevertheless, the 
numerical results presented in Fig.~\ref{fig3} show that the cross section 
$ \sigma_{{\cal VV}} (\gamma \gamma \to B_c^+ B_c^-)$ is more than an order of magnitude 
larger the section $ \sigma_{{\cal VV}} (e^+ e^- \to \gamma^\ast \to B_c^+ B_c^-) $ 
(single-photon annihilation mechanism \cite{apm1}). The cross section 
$ \sigma_{{\cal PP}} (\gamma \gamma \to B_c^+B_c^-) $ is 10 times larger than the cross section 
$ \sigma_{{\cal PP}} (e^+ e^- \to \gamma^\ast \to B_c^+B_c^-) $ \cite{apm1}.
First of all, it should be noted that a significant increase 
in the cross section for the production of a pair of vector mesons as compared 
to other mesons is also observed in $ e^+ e^- $ annihilation \cite{baranov,apm1,bbl}. The increase 
in the cross section for the production of a pair of vector mesons during the 
transition from the reaction $ e^+ e^- \to B_c^{\ast +} B_c^{\ast -} $ to 
the process $\gamma\gamma \to B_c^{\ast +} B_c^{\ast -} $ is due to the structure of total 
amplitude, which is determined by 20 Feynman diagrams. Thus, we can say 
that the observation of the pair production of vector $ B_c^\ast $ mesons in the 
$\gamma\gamma$ interaction has clear advantages. 
Our results are in agreement with
previous calculation made in \cite{baranov} in nonrelativistic approach. So, for example,
the numerical value of the production cross section $\sigma_{{\cal VV}}$ from \cite{baranov}
at the maximum point is near $10^{-7}~\mu b$ what is close to our nonrelativistic values 
in Fig.~\ref{fig3} (dashed line).
A slight difference in nonrelativistic results is related to the choice of parameters
$\Psi_{B_c}(0)$ and $\alpha_s$.
Assuming that the luminosity of the photon collider will be at least half the luminosity 
of the electron-positron collider, we can estimate the yield of pairs of vector $B_c$
mesons at 250 events per month.

In conclusion, we list the main theoretical uncertainties of our calculation. 
The calculation of all 
the main parameters that determine the cross sections for the production of a pair of 
$B_c$ mesons is performed within the framework of the quark model 
(see also \cite{apm1,apm2}). Moreover, we 
did not take into account the contribution from the parameters $\omega_{20}$, 
$\omega_{02}$, $\omega_{11}$, which in principle can be taken into account. 
They are related to corrections of order ${\bf p}^4/m_{1,2}^4$, ${\bf q}^4 /m_{1,2}^4$. 
Their contribution to the production cross section can be about $30\%$. Another important 
source of theoretical error is related to the determination of the wave function 
of the bound states of quarks in the region of relativistic momenta. Taking into 
account the contribution of the relativistic momentum region to the total value 
of parameters \eqref{eq29}, we can say that the possible theoretical error in the cross 
section does not exceed $20\%$ (the error in the wave function is estimated at $5\%$). 
Also, in our work, the radiation corrections of order $O(\alpha_s)$, 
which can be of order $20\%$ in the pair production cross section, 
were not taken into account. As a result, we estimate the total theoretical uncertainty 
of our calculations at approximately $40\%$. When our work was already published 
in the electronic archive, a work \cite{qiao} appeared in which one-loop corrections 
to the pair production cross section were calculated.

\acknowledgments
The authors are grateful to A.V. Berezhnoy for useful discussions.
The work of F.~A.~Martynenko is supported by the Foundation for the
Advancement of Theoretical Physics and Mathematics "BASIS" (grant No. 19-1-5-67-1).
\appendix
\section{The Feynman amplitudes of a production of heavy $b,c$ quarks 
and $\bar b,\bar c$ antiquarks in $\gamma\gamma$ interaction}
\begin{figure}[htbp]
\centering
\includegraphics[scale=0.95]{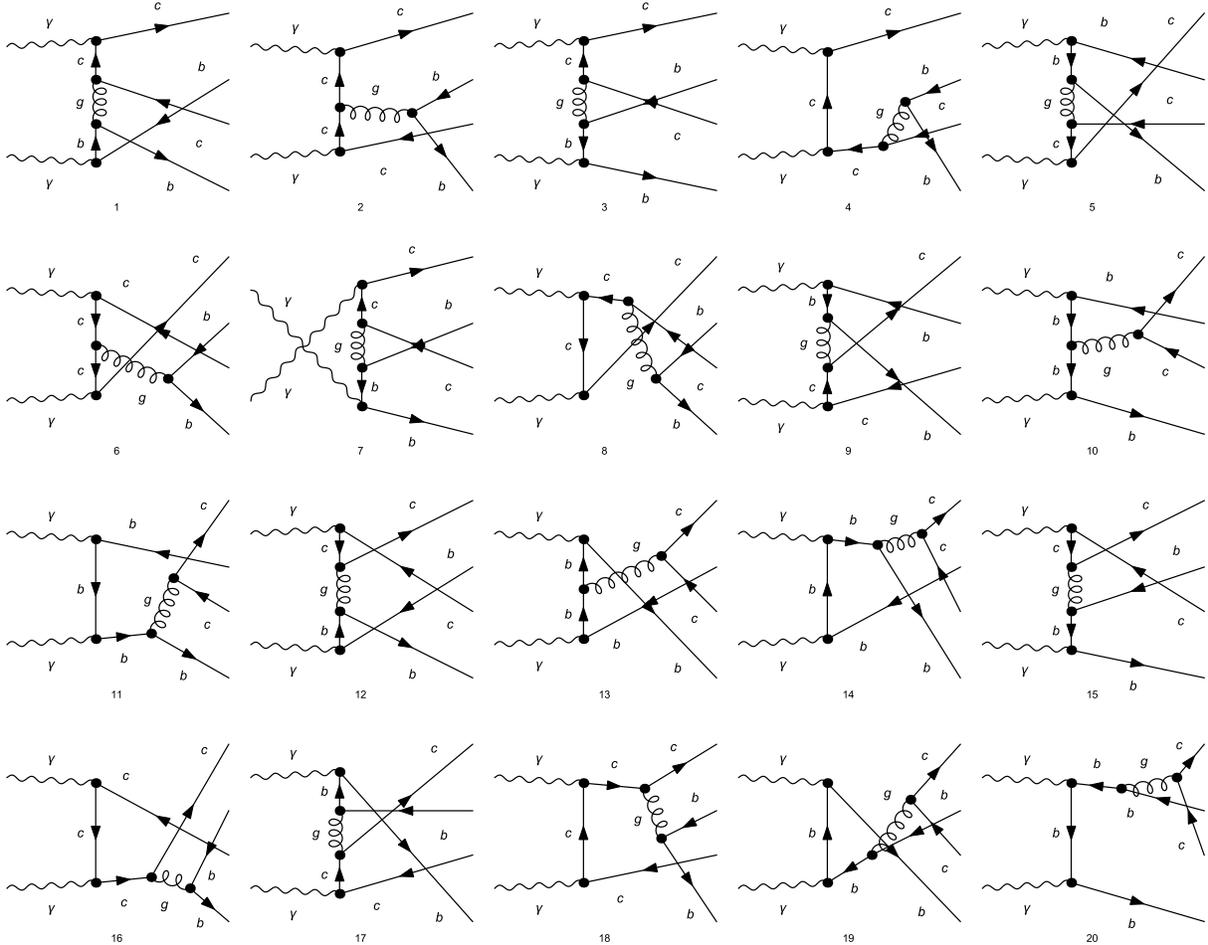}
\caption{The production of two free quarks (b,c) and antiquarks $(\bar b,\bar c)$ in 
$\gamma\gamma$ interaction in FeynArts package.
Wavy lines show the real photons.}
\label{fig4}
\end{figure}

At the first perturbative stage of the process of production of a $B_c$ meson pair 
that we are studying, we have the production of two heavy quarks $(b,c)$ 
and two heavy antiquarks $(\bar b,\bar c)$
in the photon-photon interaction. In the leading order in $ \alpha_s $, this stage 
is described by 20 Feynman amplitudes in Fig.~\ref{fig4}, which can be generated 
in the FeynArts package \cite{feynarts}.
At the second stage of the production process, quarks and antiquarks unite 
with some probability into $B_c$ mesons. The part of the Feynman amplitudes in this stage
is presented in Fig.~\ref{fig1}.

\newpage

\section{The coefficient functions $A_i^{\mu\nu}$ entering in
the pseudoscalar $B_c$ meson production amplitude (19)}
\vspace{1mm}
{\underline {$\gamma+\gamma\to B_c^++B_c^{-}$}.

\begin{equation}
A^{\mu\nu}_i=f_{i,1} v_1^\mu v_1^\nu+f_{i,2} g^{\mu\nu},
\end{equation}

\begin{equation}
f_{1,1}=
c_1 \bigl(s^2 \bigl(q_b^2 \bigl(-\frac{64 r_1^4}{3}-\frac{128 r_1^3}{3}\bigr)+q_c^2 \bigl(-\frac{128
   r_1^4}{3}+\frac{128 r_1^3}{3}+\frac{128 r_1^2}{3}-\frac{128 r_1}{3}\bigr)\bigr)-\frac{64}{3} q_b^2
   r_1^4+
\end{equation}
\begin{displaymath}
q_c^2 \bigl(512 r_1^4-1664 r_1^3+1920 r_1^2-896 r_1+128\bigr)\bigr)+c_2 \bigl(s^2
   \bigl(q_b^2 \bigl(-\frac{128 r_1^4}{3}+128 r_1^3-\frac{256 r_1^2}{3}\bigr)+
\end{displaymath}
\begin{displaymath}
q_c^2 \bigl(-\frac{64
   r_1^4}{3}+128 r_1^3-256 r_1^2+\frac{640 r_1}{3}-64\bigr)\bigr)+q_b^2 \bigl(512 r_1^4-384
   r_1^3\bigr)+q_c^2 \bigl(-\frac{64 r_1^4}{3}+\frac{256 r_1^3}{3}-128 r_1^2+
\end{displaymath}
\begin{displaymath}
\frac{256
   r_1}{3}-\frac{64}{3}\bigr)\bigr)+s^2 \bigl(q_b^2 \bigl(32 r_1^3-32 r_1^4\bigr)+q_c^2 \bigl(-32
   r_1^4+96 r_1^3-96 r_1^2+32 r_1\bigr)\bigr)-64 q_b^2 r_1^4+q_c^2 \bigl(-64 r_1^4+
\end{displaymath}
\begin{displaymath}
256r_1^3-384 r_1^2+256 r_1-64\bigr),
\end{displaymath}
\begin{equation}
f_{1,2}=
\sqrt{c_1} \sqrt{c_2} \bigl(|{\bf P}| s^3 z \bigl(16 q_b^2 r_1^2+q_c^2 \bigl(16 r_1^2-32
   r_1+16\bigr)\bigr)+s^4 \bigl(q_c^2 \bigl(-8 r_1^2+16 r_1-8\bigr)-8 q_b^2 r_1^2\bigr)+
\end{equation}
\begin{displaymath}
s^2 \bigl(48
   q_b^2 r_1^2+q_c^2 \bigl(48 r_1^2-96 r_1+48\bigr)\bigr)\bigr)+c_1 \bigl(|{\bf P}| z \bigl(s^3
   \bigl(q_b^2 \bigl(16 r_1^2-\frac{32 r_1^3}{3}\bigr)+q_c^2 \bigl(-\frac{16 r_1^3}{3}+\frac{32
   r_1^2}{3}-\frac{16 r_1}{3}\bigr)\bigr)+
\end{displaymath}
\begin{displaymath}
s \bigl(q_c^2 \bigl(128 r_1^3-288 r_1^2+192 r_1-32\bigr)-16
   q_b^2 r_1^3\bigr)\bigr)+s^4 \bigl(q_b^2 \bigl(\frac{16 r_1^3}{3}-8 r_1^2\bigr)+q_c^2 \bigl(\frac{8
   r_1^3}{3}-\frac{16 r_1^2}{3}+\frac{8 r_1}{3}\bigr)\bigr)+
\end{displaymath}
\begin{displaymath}   
s^2 \bigl(q_b^2 \bigl(\frac{32 r_1^4}{3}+\frac{56
   r_1^3}{3}-16 r_1^2\bigr)+q_c^2 \bigl(-\frac{256 r_1^3}{3}+\frac{512 r_1^2}{3}-\frac{256
   r_1}{3}\bigr)\bigr)+\frac{64 q_b^2 r_1^4}{3}\bigr)+c_2 \bigl(|{\bf P}| z \bigl(s^3 \bigl(q_b^2
   \bigl(\frac{16 r_1^3}{3}-\frac{16 r_1^2}{3}\bigr)
\end{displaymath}
\begin{displaymath}   
+q_c^2 \bigl(\frac{32 r_1^3}{3}-16
   r_1^2+\frac{16}{3}\bigr)\bigr)+s \bigl(q_b^2 \bigl(96 r_1^2-128 r_1^3\bigr)+q_c^2 \bigl(16 r_1^3-48
   r_1^2+48 r_1-16\bigr)\bigr)\bigr)+s^4 \bigl(q_b^2 \bigl(\frac{8 r_1^2}{3}-\frac{8
   r_1^3}{3}\bigr)
\end{displaymath}
\begin{displaymath}
+q_c^2 \bigl(-\frac{16 r_1^3}{3}+8 r_1^2-\frac{8}{3}\bigr)\bigr)+s^2 \bigl(q_b^2
   \bigl(\frac{256 r_1^3}{3}-\frac{256 r_1^2}{3}\bigr)+q_c^2 \bigl(\frac{32 r_1^4}{3}-\frac{184 r_1^3}{3}+104
   r_1^2-\frac{200 r_1}{3}+\frac{40}{3}\bigr)\bigr)+
\end{displaymath}
\begin{displaymath}
q_c^2 \bigl(\frac{64 r_1^4}{3}-\frac{256 r_1^3}{3}+128
   r_1^2-\frac{256 r_1}{3}+\frac{64}{3}\bigr)\bigr)+|{\bf P}| z \bigl(s^3 \bigl(q_b^2 \bigl(8 r_1^3-8
   r_1^2\bigr)+q_c^2 \bigl(-8 r_1^3+16 r_1^2-8 r_1\bigr)\bigr)+
\end{displaymath}
\begin{displaymath}
s \bigl(16 q_b^2 r_1^3+q_c^2
   \bigl(-16 r_1^3+48 r_1^2-48 r_1+16\bigr)\bigr)\bigr)+s^4 \bigl(q_b^2 \bigl(4 r_1^2-4
   r_1^3\bigr)+q_c^2 \bigl(4 r_1^3-8 r_1^2+4 r_1\bigr)\bigr)+
\end{displaymath}
\begin{displaymath}
s^2 \bigl(q_b^2 \bigl(8 r_1^2-8
   r_1^3\bigr)+q_c^2 \bigl(8 r_1^3-16 r_1^2+8 r_1\bigr)\bigr),
\end{displaymath}
\begin{equation}
f_{2,1}=
\sqrt{c_1} \sqrt{c_2} s^2 \bigl(-16 q_b^2-16 q_c^2\bigr)+c_1 \bigl(s^2 \bigl(q_b^2 \bigl(-\frac{64
   r_1^2}{3}+\frac{64 r_1}{3}-16\bigr)+q_b q_c \bigl(\frac{32 r_1^2}{3}+\frac{32
   r_1}{3}-32\bigr)
\end{equation}
\begin{displaymath}
+q_c^2 \bigl(32 r_1^2+\frac{64 r_1}{3}-16\bigr)\bigr)-32 q_b^2 r_1^2+q_b
   q_c \bigl(64 r_1-256 r_1^2\bigr)+q_c^2 \bigl(-224 r_1^2+320 r_1-64\bigr)\bigr)+
\end{displaymath}
\begin{displaymath}
c_2 \bigl(s^2
   \bigl(q_b^2 \bigl(32 r_1^2-\frac{256 r_1}{3}+\frac{112}{3}\bigr)+q_b q_c \bigl(\frac{32 r_1^2}{3}-32
   r_1-\frac{32}{3}\bigr)+q_c^2 \bigl(-\frac{64 r_1^2}{3}+\frac{64 r_1}{3}-16\bigr)\bigr)+
\end{displaymath}
\begin{displaymath}   
q_b^2 \bigl(-224
   r_1^2+128 r_1+32\bigr)+q_b q_c \bigl(-256 r_1^2+448 r_1-192\bigr)+q_c^2 \bigl(-32 r_1^2+64
   r_1-32\bigr)\bigr)+
\end{displaymath}
\begin{displaymath}
s^2 \bigl(q_b^2 \bigl(16 r_1^2-
16 r_1+8\bigr)+q_b q_c \bigl(32 r_1^2-32
r_1+16\bigr)+q_c^2 \bigl(16 r_1^2-16 r_1+8\bigr)\bigr)+
\end{displaymath}
\begin{displaymath}   
32 q_b^2 r_1^2+q_b q_c \bigl(64r_1^2-64 r_1\bigr)+
q_c^2 \bigl(32 r_1^2-64 r_1+32\bigr),
\end{displaymath}
\begin{equation}
f_{2,2}=
\sqrt{c_1} \sqrt{c_2} \bigl(|{\bf P}|^2 s^2 z^2 \bigl(8 q_b^2+8 q_c^2\bigr)+16 |{\bf P}| 
q_b q_c s^3 z+s^4\bigl(-2 q_b^2-8 q_b q_c-2 q_c^2\bigr)+
\end{equation}
\begin{displaymath}
s^2 \bigl(8 q_b^2+16 q_b q_c+8
   q_c^2\bigr)\bigr)+c_1 \bigl(|{\bf P}|^2 s^2 z^2 \bigl(8 q_b^2+8 q_c^2\bigr)+|{\bf P}| z \bigl(s \bigl(\frac{16
   q_b^2 r_1}{3}-16 q_b q_c r_1+\frac{32 q_c^2 r_1}{3}\bigr)+
\end{displaymath}
\begin{displaymath}
16 q_b q_c s^3\bigr)+s^2
   \bigl(q_b^2 \bigl(8-\frac{8 r_1}{3}\bigr)+q_b q_c (8 r_1+16)+q_c^2 \bigl(8-\frac{16
   r_1}{3}\bigr)\bigr)+s^4 \bigl(-2 q_b^2-8 q_b q_c-2 q_c^2\bigr)\bigr)+
\end{displaymath}
\begin{displaymath}
c_2 \bigl(|{\bf P}|^2 s^2 z^2
   \bigl(8 q_b^2+8 q_c^2\bigr)+|{\bf P}| z \bigl(s \bigl(q_b^2 \bigl(\frac{32}{3}-\frac{32 r_1}{3}\bigr)+q_b
   q_c (16 r_1-16)+q_c^2 \bigl(\frac{16}{3}-\frac{16 r_1}{3}\bigr)\bigr)+16 q_b q_c s^3\bigr)+
\end{displaymath}
\begin{displaymath}
s^2\bigl(q_b^2 \bigl(\frac{16 r_1}{3}+\frac{8}{3}\bigr)+q_b q_c (24-8 r_1)+q_c^2 \bigl(\frac{8
r_1}{3}+\frac{16}{3}\bigr)\bigr)+s^4 \bigl(-2 q_b^2-8 q_b q_c-2 q_c^2\bigr)\bigr)-
|{\bf P}|^2 s^2 z^2\bigl(4 q_b^2+4 q_c^2\bigr)-
\end{displaymath}
\begin{displaymath}
8 |{\bf P}| q_b q_c s^3 z+s^4 \bigl(q_b^2+4 q_b
q_c+q_c^2\bigr)+s^2 \bigl(-4 q_b^2-8 q_b q_c-4 q_c^2\bigr),
\end{displaymath}
\begin{equation}
f_{3,1}=
c_1 \bigl(s^2 \bigl(q_b^2 \bigl(-\frac{64 r_1^4}{3}-\frac{128 r_1^3}{3}\bigr)+q_c^2 \bigl(-\frac{128
   r_1^4}{3}+\frac{128 r_1^3}{3}+\frac{128 r_1^2}{3}-\frac{128 r_1}{3}\bigr)\bigr)-\frac{64}{3} q_b^2
   r_1^4+
\end{equation}
\begin{displaymath}
q_c^2 \bigl(512 r_1^4-1664 r_1^3+1920 r_1^2-896 r_1+128\bigr)\bigr)+c_2 \bigl(s^2
   \bigl(q_b^2 \bigl(-\frac{128 r_1^4}{3}+128 r_1^3-\frac{256 r_1^2}{3}\bigr)+
\end{displaymath}
\begin{displaymath}   
q_c^2 \bigl(-\frac{64
   r_1^4}{3}+128 r_1^3-256 r_1^2+\frac{640 r_1}{3}-64\bigr)\bigr)+q_b^2 \bigl(512 r_1^4-384
   r_1^3\bigr)+q_c^2 \bigl(-\frac{64 r_1^4}{3}+\frac{256 r_1^3}{3}-128 r_1^2+
\end{displaymath}
\begin{displaymath}
\frac{256
   r_1}{3}-\frac{64}{3}\bigr)\bigr)+s^2 \bigl(q_b^2 \bigl(32 r_1^3-32 r_1^4\bigr)+q_c^2 \bigl(-32
   r_1^4+96 r_1^3-96 r_1^2+32 r_1\bigr)\bigr)-64 q_b^2 r_1^4+
\end{displaymath}
\begin{displaymath}
q_c^2 \bigl(-64 r_1^4+256r_1^3-384 r_1^2+256 r_1-64\bigr),
\end{displaymath}
\begin{equation}
f_{3,2}=
\sqrt{c_1} \sqrt{c_2} \bigl(|{\bf P}| s^3 z \bigl(q_c^2 \bigl(-16 r_1^2+32 r_1-16\bigr)-16 q_b^2
   r_1^2\bigr)+s^4 \bigl(q_c^2 \bigl(-8 r_1^2+16 r_1-8\bigr)-8 q_b^2 r_1^2\bigr)+
\end{equation}
\begin{displaymath}
s^2 \bigl(48 q_b^2
   r_1^2+q_c^2 \bigl(48 r_1^2-96 r_1+48\bigr)\bigr)\bigr)+c_1 \bigl(|{\bf P}| z \bigl(s^3 \bigl(q_b^2
   \bigl(\frac{32 r_1^3}{3}-16 r_1^2\bigr)+q_c^2 \bigl(\frac{16 r_1^3}{3}-\frac{32 r_1^2}{3}+\frac{16
   r_1}{3}\bigr)\bigr)+
\end{displaymath}
\begin{displaymath}
s \bigl(16 q_b^2 r_1^3+q_c^2 \bigl(-128 r_1^3+288 r_1^2-192
   r_1+32\bigr)\bigr)\bigr)+s^4 \bigl(q_b^2 \bigl(\frac{16 r_1^3}{3}-8 r_1^2\bigr)+q_c^2 \bigl(\frac{8
   r_1^3}{3}-\frac{16 r_1^2}{3}+\frac{8 r_1}{3}\bigr)\bigr)+
\end{displaymath}
\begin{displaymath}
s^2 \bigl(q_b^2 \bigl(\frac{32 r_1^4}{3}+\frac{56
   r_1^3}{3}-16 r_1^2\bigr)+q_c^2 \bigl(-\frac{256 r_1^3}{3}+\frac{512 r_1^2}{3}-\frac{256
   r_1}{3}\bigr)\bigr)+\frac{64 q_b^2 r_1^4}{3}\bigr)+c_2 \bigl(|{\bf P}| z \bigl(s^3 \bigl(q_b^2
   \bigl(\frac{16 r_1^2}{3}-
\end{displaymath}
\begin{displaymath}
\frac{16 r_1^3}{3}\bigr)+q_c^2 \bigl(-\frac{32 r_1^3}{3}+16
   r_1^2-\frac{16}{3}\bigr)\bigr)+s \bigl(q_b^2 \bigl(128 r_1^3-96 r_1^2\bigr)+q_c^2 \bigl(-16 r_1^3+48
   r_1^2-48 r_1+16\bigr)\bigr)\bigr)+
\end{displaymath}
\begin{displaymath}
s^4 \bigl(q_b^2 \bigl(\frac{8 r_1^2}{3}-\frac{8
   r_1^3}{3}\bigr)+q_c^2 \bigl(-\frac{16 r_1^3}{3}+8 r_1^2-\frac{8}{3}\bigr)\bigr)+s^2 \bigl(q_b^2
   \bigl(\frac{256 r_1^3}{3}-\frac{256 r_1^2}{3}\bigr)+q_c^2 \bigl(\frac{32 r_1^4}{3}-\frac{184 r_1^3}{3}+104
   r_1^2-
\end{displaymath}
\begin{displaymath}
\frac{200 r_1}{3}+\frac{40}{3}\bigr)\bigr)+q_c^2 \bigl(\frac{64 r_1^4}{3}-\frac{256 r_1^3}{3}+128
   r_1^2-\frac{256 r_1}{3}+\frac{64}{3}\bigr)\bigr)+|{\bf P}| z \bigl(s^3 \bigl(q_b^2 \bigl(8 r_1^2-8
   r_1^3\bigr)+q_c^2 \bigl(8 r_1^3-16 r_1^2+
\end{displaymath}
\begin{displaymath}
8 r_1\bigr)\bigr)+s \bigl(q_c^2 \bigl(16 r_1^3-48
   r_1^2+48 r_1-16\bigr)-16 q_b^2 r_1^3\bigr)\bigr)+s^4 \bigl(q_b^2 \bigl(4 r_1^2-4
   r_1^3\bigr)+q_c^2 \bigl(4 r_1^3-8 r_1^2+4 r_1\bigr)\bigr)+
\end{displaymath}
\begin{displaymath}
s^2 \bigl(q_b^2 \bigl(8 r_1^2-8
   r_1^3\bigr)+q_c^2 \bigl(8 r_1^3-16 r_1^2+8 r_1\bigr)\bigr),
\end{displaymath}
\begin{equation}
f_{4,1}=
\sqrt{c_1} \sqrt{c_2} s^2 \bigl(-16 q_b^2-16 q_c^2\bigr)+c_1 \bigl(s^2 \bigl(q_b^2 \bigl(-\frac{64
   r_1^2}{3}+\frac{64 r_1}{3}-16\bigr)+q_b q_c \bigl(\frac{32 r_1^2}{3}+\frac{32
   r_1}{3}-32\bigr)+
\end{equation}
\begin{displaymath}
q_c^2 \bigl(32 r_1^2+\frac{64 r_1}{3}-16\bigr)\bigr)-32 q_b^2 r_1^2+q_b
   q_c \bigl(64 r_1-256 r_1^2\bigr)+q_c^2 \bigl(-224 r_1^2+320 r_1-64\bigr)\bigr)+
\end{displaymath}
\begin{displaymath}
c_2 \bigl(s^2
   \bigl(q_b^2 \bigl(32 r_1^2-\frac{256 r_1}{3}+\frac{112}{3}\bigr)+q_b q_c \bigl(\frac{32 r_1^2}{3}-32
   r_1-\frac{32}{3}\bigr)+q_c^2 \bigl(-\frac{64 r_1^2}{3}+\frac{64 r_1}{3}-16\bigr)\bigr)+
\end{displaymath}
\begin{displaymath}
q_b^2 \bigl(-224
   r_1^2+128 r_1+32\bigr)+q_b q_c \bigl(-256 r_1^2+448 r_1-192\bigr)+q_c^2 \bigl(-32 r_1^2+64
   r_1-32\bigr)\bigr)+
\end{displaymath}
\begin{displaymath}
s^2 \bigl(q_b^2 \bigl(16 r_1^2-16 r_1+8\bigr)+q_b q_c \bigl(32 r_1^2-32
   r_1+16\bigr)+q_c^2 \bigl(16 r_1^2-16 r_1+8\bigr)\bigr)+32 q_b^2 r_1^2+
\end{displaymath}
\begin{displaymath}
q_b q_c \bigl(64r_1^2-64 r_1\bigr)+q_c^2 \bigl(32 r_1^2-64 r_1+32\bigr),
\end{displaymath}
\begin{equation}
f_{4,2}=
\sqrt{c_1} \sqrt{c_2} \bigl(|{\bf P}|^2 s^2 z^2 \bigl(8 q_b^2+8 q_c^2\bigr)-16 |{\bf P}| q_b q_c s^3 z+s^4
   \bigl(-2 q_b^2-8 q_b q_c-2 q_c^2\bigr)+
\end{equation}
\begin{displaymath}
s^2 \bigl(8 q_b^2+16 q_b q_c+8
   q_c^2\bigr)\bigr)+c_1 \bigl(|{\bf P}|^2 s^2 z^2 \bigl(8 q_b^2+8 q_c^2\bigr)+|{\bf P}| z \bigl(s \bigl(-\frac{16
   q_b^2 r_1}{3}+16 q_b q_c r_1-\frac{32 q_c^2 r_1}{3}\bigr)-
\end{displaymath}
\begin{displaymath}
16 q_b q_c s^3\bigr)+s^2
\bigl(q_b^2 \bigl(8-\frac{8 r_1}{3}\bigr)+q_b q_c (8 r_1+16)+q_c^2 \bigl(8-\frac{16
r_1}{3}\bigr)\bigr)+s^4 \bigl(-2 q_b^2-8 q_b q_c-2 q_c^2\bigr)\bigr)+
\end{displaymath}
\begin{displaymath}
c_2 \bigl(|{\bf P}|^2 s^2 z^2
\bigl(8 q_b^2+8 q_c^2\bigr)+|{\bf P}| z \bigl(s \bigl(q_b^2 \bigl(\frac{32 r_1}{3}-
\frac{32}{3}\bigr)+q_b
q_c (16-16 r_1)+q_c^2 \bigl(\frac{16 r_1}{3}-\frac{16}{3}\bigr)\bigr)-16 q_b q_c s^3\bigr)
\end{displaymath}
\begin{displaymath}
+s^2\bigl(q_b^2 \bigl(\frac{16 r_1}{3}+\frac{8}{3}\bigr)+q_b q_c (24-8 r_1)+q_c^2 \bigl(\frac{8
r_1}{3}+\frac{16}{3}\bigr)\bigr)+s^4 \bigl(-2 q_b^2-8 q_b q_c-2 q_c^2\bigr)\bigr)+
\end{displaymath}
\begin{displaymath}
|{\bf P}|^2 s^2 z^2\bigl(-4 q_b^2-4 q_c^2\bigr)
+8 |{\bf P}| q_b q_c s^3 z+s^4 \bigl(q_b^2+4 q_b
q_c+q_c^2\bigr)+s^2 \bigl(-4 q_b^2-8 q_b q_c-4 q_c^2\bigr),
\end{displaymath}
\begin{equation}
f_{5,1}=
-64 \sqrt{c_1} \sqrt{c_2} q_b q_c s^2+c_1 \bigl(q_b q_c \bigl(-64 r_1^2-\frac{64
   r_1}{3}\bigr) s^2+q_b q_c \bigl(\frac{1280 r_1^2}{3}-128 r_1\bigr)\bigr)+
\end{equation}
\begin{displaymath}
c_2 \bigl(q_b q_c
   \bigl(-64 r_1^2+\frac{448 r_1}{3}-\frac{256}{3}\bigr) s^2+q_b q_c \bigl(\frac{1280 r_1^2}{3}-\frac{2176
   r_1}{3}+\frac{896}{3}\bigr)\bigr)+
\end{displaymath}
\begin{displaymath}
q_b q_c \bigl(64 r_1-64 r_1^2\bigr) s^2+q_b q_c \bigl(128r_1-128 r_1^2\bigr),
\end{displaymath}
\begin{equation}
f_{5,2}=
\sqrt{c_1} \sqrt{c_2} \bigl(32 |{\bf P}|^2 q_b q_c s^2 z^2-8 q_b q_c s^4+96 q_b q_c
   s^2\bigr)+c_1 \bigl(32 |{\bf P}|^2 q_b q_c s^2 z^2+
\end{equation}
\begin{displaymath}
q_b q_c \bigl(\frac{128 r_1^2}{3}-\frac{80
r_1}{3}-32\bigr) s^2+\frac{256}{3} q_b q_c r_1^2-8 q_b q_c s^4\bigr)+c_2 \bigl(32 |{\bf P}|^2
q_b q_c s^2 z^2+
\end{displaymath}
\begin{displaymath}
q_b q_c \bigl(\frac{128 r_1^2}{3}-\frac{176 r_1}{3}-16\bigr) s^2+q_b q_c
\bigl(\frac{256 r_1^2}{3}-\frac{512 r_1}{3}+\frac{256}{3}\bigr)-8 q_b q_c s^4\bigr)-
\end{displaymath}
\begin{displaymath}
16 |{\bf P}|^2 q_bq_c s^2 z^2+4 q_b q_c s^4+16 q_b q_c s^2,
\end{displaymath}

In all these equations the substitution $q_c\to Q_c\sqrt{\alpha_{s,b}}$, 
$q_b\to Q_b\sqrt{\alpha_{s,c}}$ is needed.
In these expressions, the role of relativistic corrections is played by the functions 
$c_1(q)=(\varepsilon_1(q)-m_1)/(\varepsilon_1(q)+m_1)$, 
$c_2(q)=(\varepsilon_2(q)-m_2)/(\varepsilon_2(q)+m_2)$, which are subsequently 
converted into relativistic parameters \eqref{eq29}.


\begin{thebibliography}{99}
\bibitem{gklt}S.~S.~Gershtein, V.~V.~Kiselev, A.~K.~Likhoded, and A.~V.~Tkabladze,  
Phys. Usp. {\bf 38}, 1 (1995).
\bibitem{brambilla2011}N.~Brambilla, S.~Eidelman, B.~K.~Heltsley et al. Eur. Phys. J. C 
{\bf 71}, 1534 (2011).
\bibitem{rqm1}D.~Ebert, R.~N.~Faustov and V.~O.~Galkin, Phys. Rev. D {\bf 67}, 014027 (2003).
\bibitem{pot}N.~Brambilla, A.~Pineda, J.~Soto and A.~Vairo,  Phys. Rev. D {\bf 63}, 014023 (2001).
\bibitem{eq}E.~J. Eichten and C.~Quigg, Phys. Rev. D {\bf 99}, 054025 (2019).
\bibitem{eb}E.~Braaten and J.~Lee, Phys. Rev. D {\bf 67}, 054007 (2003); Phys.
Rev. D {\bf 72}, 099901(E) (2005).
\bibitem{qiao}K.~Hagiwara, E.~Kou and C.-F.~Qiao, Phys. Lett. B {\bf 570}, 39 (2003).
\bibitem{chao1}K.-Y.~Liu, Z.-G.~He and K.-T.~Chao, Phys. Lett. B {\bf 557}, 45 (2003).
\bibitem{bodwin1}G.~T.~Bodwin, D.~Kang and J.~Lee, Phys. Rev. D {\bf 74}, 014014 (2006).
\bibitem{apm3}D.~Ebert and A.~P.~Martynenko, Phys. Rev. D {\bf 74}, 054008 (2006).
\bibitem{apm4}D.~Ebert, R.~N.~Faustov, V.~O.~Galkin and A.~P.~Martynenko, 
Phys. Lett. B {\bf 672}, 264 (2009).
\bibitem{serbo1}I.~F.~Ginzburg, G.~L.~Kotkin, V.~G.~Serbo and V.~I.~Tel'nov, 
JETP Lett. {\bf 34}, 491 (1981).
\bibitem{serbo2}E.~A.~Kuraev, A.~Schiller, and V.~G.~Serbo, Phys. Lett. {\bf 134}, 455 (1984).
\bibitem{carlo}C.~Brikporad, , R.~H. Milburn, N.~ Tanaka, and M.~Fotino, Phys. Rev. {\bf 138},
B1546 (1965).
\bibitem{telnov}V.~I.~Telnov, Nucl. Part. Phys. Proc. {\bf 273-275}, 219 (2016).
\bibitem{badelek}B.~Badelek et al. TESLA Technical Design Report, Part VI, Ch. 1, Photon
Collider at TESLA, DESY 2001-011, hep-ex/0108012.
\bibitem{akl1}A.~V.~Berezhnoy, A.~K.~Likhoded and M.V. Shevlyagin, Phys. Lett. B 
{\bf 342} 351 (1995).
\bibitem{akl2}A.~V.~Berezhnoy, V.V. Kiselev and A.~K.~Likhoded, 
Phys. Lett. B {\bf 381} 341 (1996).
\bibitem{baranov}S.~P.~Baranov, Phys. Rev. D {\bf 55}, 2756 (1997).
\bibitem{apm1}A.~A.~Karyasov, A.~P.~Martynenko and F.~A.~Martynenko, 
Nucl. Phys. B {\bf 911}, 36 (2016).
\bibitem{apm2}A.~V.~Berezhnoy, A.~P.~Martynenko and F.~A.~Martynenko and O.~S.~Sukhorukova,
Nucl. Phys. A {\bf 986}, 34 (2019).
\bibitem{pdg}M.~Tanabashi et al. (Particle Data Group), Phys. Rev. D 98, 030001 (2018).
\bibitem{aaij}R. Aaij et al. (LHCb Collaboration), Phys. Rev. Lett. {\bf 122}, 232001 (2019).
\bibitem{avb}A.~V.~Berezhnoy,  I.~N.~Belov A.~K.~Likhoded and A.~V.~Luchinsky, 
Mod. Phys. Lett. A {\bf 34} 1950331 (2019).
\bibitem{LHCb}R.~Aaij et al. (LHCb collaboration), arXiv:2004.08163v1 [hep-ex].
\bibitem{feynarts}J.~Kublbeck, M.~B\"ohm, and A.~Denner, Comp. Phys. Comm. {\bf 60},
165 (1990). 
\bibitem{feynarts1}T.~Hahn, Comp. Phys. Commun. {\bf 140}, 418 (2001).
\bibitem{apm5}E.~N.~Elekina and A.~P.~Martynenko, Phys. Rev. D {\bf 81}, 054006 (2010).
\bibitem{apm6}A.P.~Martynenko and A.M.~Trunin, Phys. Rev. D {\bf 86}, 094003 (2012).
\bibitem{brodsky}S.~J.~Brodsky and J.~R.~Primack, Ann. Phys. {\bf 52}, 315 (1969).
\bibitem{faustov}R.~N.~Faustov, Ann. Phys. {\bf 78}, 176 (1973).
\bibitem{bodwin2002}G.~T.~Bodwin and A.~Petrelli, Phys. Rev. D {\bf 66}, 094011 (2002).
\bibitem{apm8}A.~P.~Martynenko, Phys. Rev. D {\bf 72}, 074022 (2005).
\bibitem{apm9}A.P.~Martynenko and A.M.~Trunin, Eur. Phys. J. C {\bf 75}, 138 (2015). 
\bibitem{form}J.~Kuipers, T.~Ueda, J.~A.~M.~Vermaseren, and J.~Vollinga, 
Comput. Phys. Commun. {\bf 184}, 1453 (2013). 
\bibitem{repko2}S.~F.~Radford and W.~W.~Repko, Phys. Rev. D {\bf 75}, 074031 (2007).
\bibitem{repko3}S.~N.~Gupta, J.~M.~Johnson, W.~W.~Repko and C.~J.~Suchyta, 
Phys. Rev. D {\bf 49}, 1551 (1994).
\bibitem{rqm3}D.~Ebert, R.~N.~Faustov and V.~O.~Galkin, Phys. Rev. D {\bf 72}, 034026 (2005).
\bibitem{godfrey}S.~Godfrey, Phys. Rev. D {\bf 70}, 054017 (2004).
\bibitem{lucha}W.~Lucha and F.~F.~Sch\"oberl, Phys. Rev. A {\bf 51}, 4419 (1995). 
\bibitem{bbl}G.~T.~Bodwin, E.~Braaten and J.~Lee, Phys. Rev. D {\bf 67}, 054023 (2003),
Phys. Rev. D {\bf 72}, 099904 (2005) (erratum).
\bibitem{qiao}Z.-Q.~Chen, H.~Yang and C.-F.~Qiao, arXiv:2005.07317[hep-ph].
\end{thebibliography}
\end{document}